\documentclass[aps, reprint, twocolumn, superscriptaddress, citeautoscript]{revtex4-1}

\usepackage{graphicx}

\usepackage{amssymb}
\usepackage{amsmath}
\usepackage{amsfonts}
\usepackage{mathrsfs}

\usepackage{bm}
\usepackage{dcolumn}
\usepackage{color}
\usepackage[colorlinks=true,citecolor=blue]{hyperref}
\hypersetup{colorlinks=true,citecolor=blue,linkcolor=red, urlcolor=blue}

\usepackage{float}

\usepackage{color}
\usepackage{ulem}
\definecolor{purple}{rgb}{0.8,0,0.6}
\definecolor{orange}{rgb}{1,0.55,0}

\newcommand{\beqn}{\begin{eqnarray}}
	\newcommand{\eeqn}{\end{eqnarray}}

\newcommand{\comment}[1]{}

\begin{document}
	
	\title[]{Coherent Phonons-Driven Hot Carrier Effect in a Superlattice Solar Cell}
	
	\author{I. Makhfudz}
	\affiliation{IM2NP, UMR CNRS 7334, Aix-Marseille Universit\'{e}, 13013 Marseille, France}
	\author{N. Cavassilas}
	\affiliation{IM2NP, UMR CNRS 7334, Aix-Marseille Universit\'{e}, 13013 Marseille, France}
	\author{Y. Hajati}
	\affiliation{Department of Physics, Faculty of Science, Shahid Chamran University of Ahvaz, 6135743135 Ahvaz, Iran}
	\author{H. Esmaielpour}
	\affiliation{Walter Schottky Institut, Technische Universität München, Am Coulombwall 4, D-85748 Garching, Germany}
	\author{F. Michelini}
	\affiliation{IM2NP, UMR CNRS 7334, Aix-Marseille Universit\'{e}, 13013 Marseille, France}

	\date{\today}

	\begin{abstract}
		Carrier thermalization in a superlattice solar cell made of polar semiconductors is studied theoretically by considering a minimal model where electron-phonon scattering is the principal channel of carrier energy loss. Importantly, the effect of an intrinsic quantum mechanical property; the phonon coherence, on carrier thermalization is investigated, within semiclassical picture in terms of phonon wave packet. It turns out that coherent longitudinal optical (LO) phonons weaken the effective electron-phonon coupling, thus supposedly lowering the carrier energy loss rate in solar cell. The resulting thermalization power is indeed significantly reduced by the coherent phonons, resulting in enhanced hot carrier effect, particularly for thin enough well layer where carrier confinement is also strong. A recent experiment on superlattice solar cell prototype is shown to manifest the coherent phonons-driven phenomenon. Our results demonstrate the practical implications of the fundamental quantum coherence property of phonons in semiconductors for improving superlattice solar cell performance, via hot carrier effect.
	\end{abstract}
	
	\maketitle
	
	\section{Introduction}
	Superlattice has emerged as a remarkably useful semiconductor heterostructure since its conception by L. Esaki and R. Tsu \cite{EsakiIBMjournal} and exposition by L. Esaki and L. L. Chang of novel transport phenomenon in such artificially made structure, such as negative differential resistance \cite{EsakiChangPRL}. The band structure of superlattice has been studied using various methods \cite{RMPsuperlattice}, consisting of mainly mini-subbands mimicking discrete energy levels of a quantum well but each of which has a finite band width describing the tunneling of wave function overlap between states localized at neighboring quantum wells constituting the superlattice. 
	
	Arising as a historically imminent component in the study of semiconductor physics and nanoelectronics \cite{TsuBook}, superlattice also has immediate practical applications, including for photovoltaic solar cell devices. In the latter, the interest is to maximize the generated output from such solar cell device. One of the ways in which this can be achieved is by keeping the photogenerated charge carriers to remain energetic before their extraction. In other words, it is necessary to keep the charge carriers ``hot", hence the so-called hot carrier effect \cite{Conibeer1}-\cite{JShah}, emerging as a leading strategy to increase solar cell efficiency.  Recent work by us \cite{PRapplied} found that a quantum well heterostructure significantly reduces the thermalization power, equivalent to energy loss rate, of the photogenerated charge carriers, leading to hot carrier effect in such quantum well or ultra-thin solar cells \cite{PRappliedCavassilas}. 
	
	Motivated by existing works \cite{BarnhamDuggan}-\cite{EsmaielpourAPL} to provide a microscopic theory of carrier thermalization in a superlattice solar cell, we extend our study on a quantum well solar cell to a superlattice-based solar cell, focusing on the role of electron-phonon-induced scattering. Our present work in particular addresses the quantum coherence of phonons and its effect on carrier thermalization in superlattice solar cell, within the framework of semi-classical model of kinetic theory which incorporates the effects of quantum confinement, tunneling, and coherence, in a model with phonon band folding \cite{folding}\cite{Kleinreview}. The investigation of carrier thermalization is critical to the study of hot carrier effect \cite{Conibeer1}-\cite{JShah}, by which the efficiency of the solar cell can be improved \cite{RossNozik} beyond the paradigmatic Shockley-Queisser theoretical limit \cite{ShockleyQueisser}. Our theory will show, first, that phonon coherence reduces the effective electron-phonon coupling, and second, that when phonons are coherent and the carriers are highly confined, carrier thermalization is strongly suppressed, driving a significant hot carrier effect, thus offering an alternative method to enhance solar cell performance. 
	
	The paper starts with the Hamiltonian that describes the carriers (electrons), the phonons, and the interactions between them. The concept of coherent phonons is then discussed and its role on modifying the electron-phonon coupling is elucidated. The resulting electron-phonon scattering rate in the superlattice with coherent phonon effect is derived. The following section analyzes the thermalization power that measures the rate of electron energy loss via electron-phonon scattering. The theoretical prediction on thermalization power is compared with experimental data, indicating manifestation of coherent phonons-driven hot carrier effect. The paper ends with discussion and conclusion.
	
	\section{Model Hamiltonian}
	\subsection{Electron Hamiltonian}
	With a profile of superlattice as shown in Fig. 1, we consider a tight-binding Hamiltonian written in the basis of states localized in the well $|\alpha=1,2,3,\cdots\rangle$ with tunneling only between adjacent wells separated by a barrier. In this case, the Hamiltonian in the (first-quantized) matrix form can be written as
	
	\begin{equation}\label{electronHamiltonian}
		H_n=\sum_{\alpha\beta}E_n|\alpha\rangle\langle \beta|\delta_{\alpha\beta}-V_n|\alpha\rangle\langle \beta|(\delta_{\alpha,\beta-1}+\delta_{\alpha,\beta+1})
	\end{equation}
	where the $\alpha,\beta$ represent the well indices while the subscript $n=1,\cdots,N_s$ represents the $n^{\mathrm{th}}$ subband of $N_s$ confined states of carrier inside the well. The $E_n$ are the energy level of the $n^{\mathrm{th}}$ subband of the individual well which for a finite potential well is given by
	\begin{equation}
		E_n(k_{\perp})=\frac{\hbar^2k^2_{\perp}}{2m_c}+E_n(a)
	\end{equation}
	where $\hbar$ is the Dirac (reduced Planck) constant, $k_{\perp}$ is the carrier's transverse wave vector, $m_c$ is the carrier mass, $E_n(a)$ is the $n^{\mathrm{th}}$ quantized energy level of a finite potential well of width $a$ and depth $V_c$ in Fig. 1 while $V_n$ is a phenomenological tunneling integral which depends on $n$. In this work, only electron processes will be of interest and ``carrier" thus refers to electron. To be concrete, we will consider an InAs(well)/AlAsSb(barrier) superlattice, with material parameters given in \cite{SupplementaryMaterials}.  
	
	The solution of Schr\:{o}dinger equation for this Hamiltonian in terms of eigenvalues is given by 
	\begin{equation}\label{superlatticedispersion}
		E_n(\mathbf{k})=E_n(k_{\perp})-2V_n\cos k_z (a+b)
	\end{equation}
	giving rise to a ``miniband" of width $4V_n$. Noting that the band width of the $n^{\mathrm{th}}$ mini-subband is $4V_n$, in this work the magnitude of $V_n$ in Eq.(\ref{electronHamiltonian}) is to be approximated as $1/2$ of the energy splitting between two energy levels belonging to $n^{\mathrm{th}}$-subband in a numerically exact Hamiltonian diagonalization calculation of a system of two quantum wells each of thickness $a$ separated by a barrier of thickness $b$. In the energy dispersion Eq.(\ref{superlatticedispersion}), $k_z$ is the wave vector of the electron traversing the superlattice; $k_z\in [-\pi/(a+b),\pi/(a+b)]$. 
	
	\begin{figure}
		\centering
		\includegraphics[angle=0,origin=c, scale=0.28]{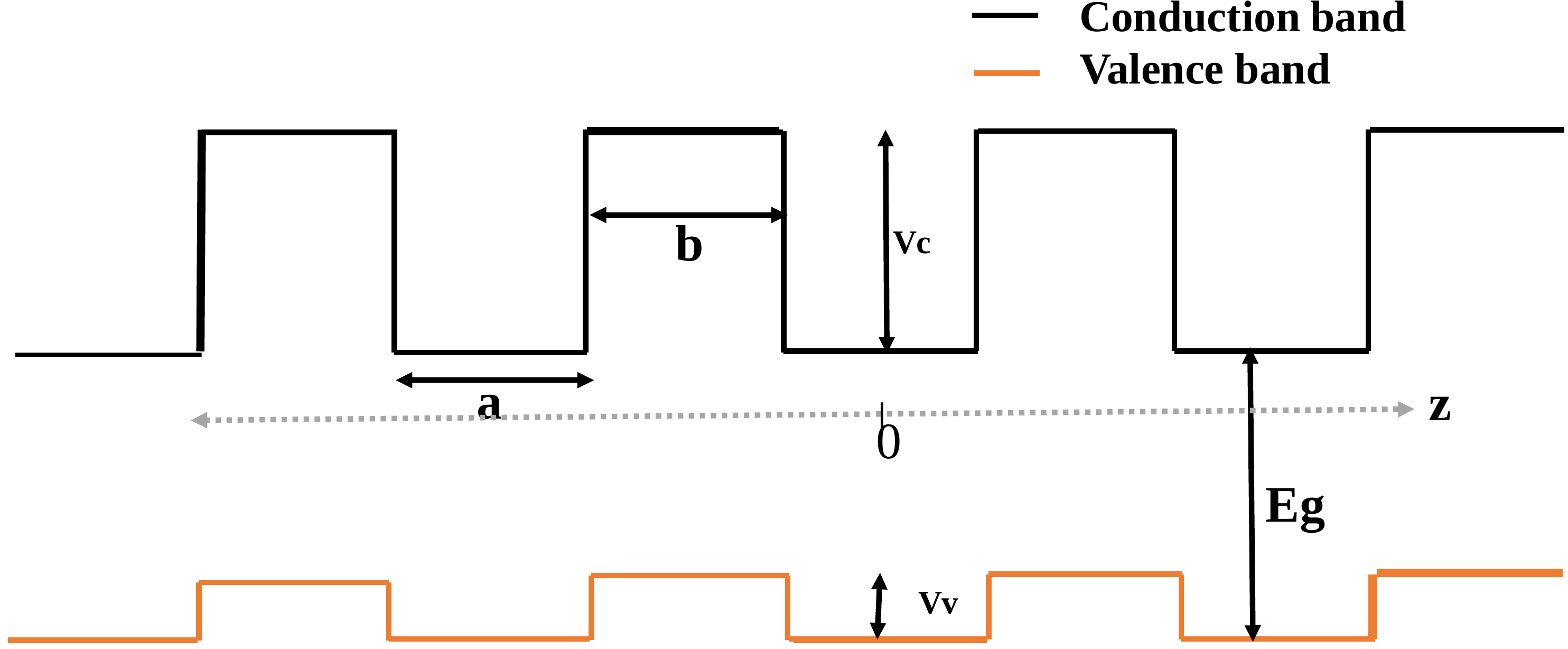}
		\label{fig:SuperlatticeGeometry}
		\caption{
			The profile of a superlattice consisting of wells of width $a$ and barriers of width $b$ with the horizontal dashed line serving as spatial coordinate axes.}
	\end{figure}
	
	The corresponding eigenfunction is
	\begin{equation}\label{electroneigenfunction}
		\psi_{n \mathbf{k}}(z)=e^{i\mathbf{k}_{\perp}\cdot\mathbf{r}_{\perp}}\frac{1}{\sqrt{N}}\sum^N_{\alpha=1}e^{ik_z\alpha(a+b)}\phi_n(z-\alpha(a+b))
	\end{equation}
	where $N$ is the number of periods (one period contains one well and one barrier) in the superlattice and $\phi_n(z)$ is the eigenfunction of the $n^{\mathrm{th}}$ subband (confined electron state) of the individual well. In the rest of this text, $n$ represents the $n^{\mathrm{th}}$ eigenstate of the electron confined in the well whereas $\alpha$ represents the $\alpha^{\mathrm{th}}$ quantum well. 
	
	\subsection{Electron-Phonon Hamiltonian}For the electron-phonon Hamiltonian, we will assume Frochlich type of Hamiltonian with 3D (bulk) phonon, with phonon band folding eventually taken into account in the calculation. In  polar semiconductors, optical phonons play the key role in carrier thermalization. In this case, the electron-phonon coupling Hamiltonian in second quantized form can be written as
	\begin{equation}\label{Frohlich}
		H_{e-phon}=\sum_{\mathbf{k},n; \mathbf{k}',n'}\sum_{\mathbf{q}}a^{\dag}_{\mathbf{k}n}a_{\mathbf{k}'n'}(b_{\mathbf{q}}-b^{\dag}_{-\mathbf{q}})M_{\mathbf{q}}G(\mathbf{k}n\mathbf{k}'n'\mathbf{q})
	\end{equation}
	where $a^{\dag}_{\mathbf{k}n}(a_{\mathbf{k}'n'})$ are the electron creation (annihilation) operator for subband $n(n')$ at wave vector $\mathbf{k}(\mathbf{k}')$ while $b^{\dag}_{-\mathbf{q}}(b_{\mathbf{q}})$ are the corresponding operators of wave vector  $\mathbf{q}$ for longitudinal optical (LO) phonon. The electron-phonon coupling function $M_{\mathbf{q}}$ is given by
	\begin{equation}
		M_{\mathbf{q}}=\sqrt{\frac{\hbar\omega_{LO} e^2}{8\pi q^2 V}\left(\frac{1}{\varepsilon_{\infty}}-\frac{1}{\varepsilon_0}\right)}
	\end{equation}
	where $\omega_{LO},e,V,\varepsilon_{\infty},\varepsilon_0$ are respectively the LO phonon frequency, elementary charge, phonon spatial volume, infinite frequency and static dielectric constants, while $G(\mathbf{k}n\mathbf{k}'n'\mathbf{q})$ is the overlap integral given by
	\begin{equation}\label{overlapfactor}
		G(\mathbf{k}n\mathbf{k}'n'\mathbf{q})=\int d^3r \psi^*_{n\mathbf{k}}(\mathbf{r}_{\perp},z)e^{i(q_z z+\mathbf{q}_{\perp}\cdot\mathbf{r}_{\perp})} \psi_{n'\mathbf{k}'}(\mathbf{r}_{\perp},z)
	\end{equation}
	where, in conjunction with Eq.(\ref{electroneigenfunction}), we will use
	\begin{equation}\label{eigenfunction}
		\phi_n(z)=\sqrt{\frac{2}{a}}\cos\frac{n\pi z}{a}
	\end{equation}
	from the eigenstate of an infinite potential well, as an analytical wave function approximating the actual wave function of the confined electron in the well layer. The detailed analytical expression for the resulting overlap integral form factor is given in \cite{SupplementaryMaterials}.
	
	In semiconductor superlattice, linear-chain model calculation shows that optical phonon band structure is folded into a smaller Brillouin zone, corresponding to confinement of the optical phonons into the well layer \cite{foldingS}. The resulting band dispersion within this mini Brillouin zone consists of zig-zag dispersion for the acoustic phonon and a close ensemble of flat bands for optical phonon \cite{foldingS}\cite{KleinreviewS}. The energy dispersion of the longitudinal optical phonon becomes qualitatively the same as that of the bulk phonon; the only difference being that the Brillouin zone is folded from $|\mathbf{q}|\in[0,\pi/a_{\mathrm{lattice}}]$ for bulk optical phonon to $|\mathbf{q}|\in[0,\pi/(a+b)]$ for optical phonon in superlattice. In a semiclassical description for superlattice, one can thus use the plane-wave function $\exp(iq_zz)$ but with $q_z\in[0,\pi/(a+b)]$, as implemented in the present work.
	
	\subsection{Phonon Hamiltonian}The phonon Hamiltonian consists of the non-interacting phonon Hamiltonian in terms of harmonic oscillators. In addition, our model also includes phonon anharmonicity \cite{RidleyBook}\cite{Klemens}, that is, phonon-phonon interaction describing the transformation of an LO phonon to two acoustic phonons (1 and 2)\cite{RidleyBook}. The acoustic phonons are assumed to be always in equilibrium at lattice temperature $T_L$ while the LO phonons may go out of equilibrium, corresponding to the non-equilibrium hot phonon effect, that may play a key role in solar cell physics \cite{Tsai}. The simplest model is employed; the phonon-phonon scattering is described within relaxation time approximation where the  scattering time is assumed to be a constant independent of wave vector 
	\begin{equation}\label{LOPacousticscatteringrate}
		\tau^{LO-ac}_{\mathbf{q}}=\frac{\tau^{LO-ac}_0}{N_{1}+N_{2}+1}
	\end{equation}
	where $\tau^{LO-ac}_0$ is the $T=0$K value of $\tau^{LO-ac}_{\mathbf{q}}$, $N_{1(2)}=(\exp(E_{1(2)}/(k_BT_L))-1)^{-1}$ of the two acoustic phonons of energies $E_1$ and $E_2$, with $k_B$ the Boltzmann constant, and the lattice temperature $T_L$. This description for phonon-phonon scattering time is assumed to hold even with phonon band folding, justified in a model where phonon-phonon coupling depends only on the energies of the involved phonons \cite{Klemens} with equal partition of the energy of an LO phonon to the two acoustic phonons, that is, $E_1=E_2=E_0/2$ where $E_0=\hbar\omega_{LO}$ is the LO phonon energy. The optical phonons are effectively confined or quasi-confined in the wells, because the optical phonons encounter and sense the abrupt change in the degree of polarizability at the boundary between the two polar materials, which effectively serves as a confining wall for the optical phonons. The existence of such confined or quasi-confined optical phonon modes is well established in literature, even for a superlattice made of similar materials (for example, two polar semiconductors of III-V family, as considered in this work) due to the sensitivity of the optical phonons to the difference in materials' polarizabilities, leading to quasi-confined optical phonon modes \cite{StroscioDutta}. As a final note, while the electron Hamiltonian, electron-phonon Hamiltonian, and phonon-phonon interaction are respectively written in first quantized, second-quantized and phenomenological forms, their treatment is unified in the semiclassical theory that is developed in this work. 
	
	\section{Coherent Phonons}Phonons are quantum particle excitations of lattice vibrations. As such, phonons posses some intrinsically quantum mechanical properties, such as bosonic quantum statistics and can also be in coherent state. Phonons are coherent when there is a well-defined phase relationship between vibrations of atoms separated by distance shorter than or equal to the phonon coherence length $l_c$ \cite{VolzPRB}. The study of phonon coherence on quantum transport in semiconductor heterostructures has remained important research problem over decades to recent years \cite{MahanPRL}-\cite{KomirenkoPRB}.
	
	In characterizing the coherence property of phonons, one can speak of spatial coherence and temporal coherence. The spatial coherence of phonon can be described in terms of phonon wave packet. Indeed, intuitively speaking, coherence can be associated with a wave-like picture of phonons, just like the wave function description of quantum mechanics. Conversely, decoherence which would give classical picture of phonons, can be associated with description of phonons as (classical) particle. The atomic vibrations have well-defined phase relationship (that is, coherent) when they are located within the envelope of the wave packet peak. Outside this envelope, their vibrations are random, and their average displacement (averaged over the atoms) vanish. The semiclassical framework in terms of wave packet describes the phonon in terms of the displacement function of the atoms in a one-dimensional atomic chain treatment of phonons, rather than in terms of the eigenfunctions of quantized Hamiltonian for phonons. 
	
	In addition to spatial coherence, in general, phonon also has temporal coherence. In the literature, the phonon temporal coherence has been discussed in terms of fully quantum mechanical description involving eigenstates of harmonic oscillator model of phonon and the so-called phonon coherent states \cite{StantonPRL}. The phonon temporal coherence is however less relevant in the present work because we assume Frohlich type of electron-phonon coupling Hamiltonian that does not depend on time. In addition, our theory eventually pertains only to steady-state situation of carrier thermalization process where no explicit time dependence shows up. Furthermore, as shown in \cite{LatourChalopinPRB}, phonon temporal coherence corresponds to its mean free path. Longitudinal optical (LO) phonons on the other hand have nearly-flat energy dispersion. In fact, in our calculation, we have approximated the LO phonons dispersion to be a constant, wave vector-independent energy dispersion $\hbar\omega_{\mathrm{LO}}=30$ meV. As such, the effective group velocity of LO phonons is zero; they do not propagate. As a result, in this description, the LO phonons do not propagate and their mean-free path vanishes. Therefore, the phonon temporal coherence length is irrelevant in this case; the LO phonons are completely incoherent in time. 
	
	In photovoltaic research, the importance of this fundamental property of phonon coherence has not been explored, despite the crucial role of phonons on carrier energy loss in a solar cell. The question that we aim to answer is the following: What is the effect of phonon coherence on carrier thermalization via electron-phonon scattering? To this end, we develop a semiclassical picture in terms of phonon wave packet where the atoms vibrate in phase when they are within the length scale (the ``width") of the wave packet relative to the center of the latter, which defines $l_c$
	as the phonon coherence length. In the present work, only coherence of LO phonons is considered; acoustic phonons are assumed to remain incoherent. Furthermore, only spatial coherence is relevant in this case, but the temporal coherence is not, because while the former corresponds to spatial localization, the latter corresponds to mean free path \cite{LatourChalopinPRB}, which vanishes for optical phonons because they do not propagate. 
	
	For simplicity, we will assume a one-dimensional (1D) wave packet along $z$, described in terms of the displacement function in a 1D (atom chain) model in the classical treatment of phonon, as given by
	\begin{equation}
		u'_{q_z}(z,t) = W'_{q_z} e^{i(q_z z-\omega_{q_z} t)}
	\end{equation}
	where $W'_{q_z}$ is the weight function of the mode $q_z$ of the phonon. This weight $W'_{q_z}$ depends on the initial condition $u'(z,0)=(1/\sqrt{\pi}q_{UV})\int^{q_{UV}}_0dq_zu'_{q_z}(z,0)$ where $q_{UV}=\pi/(a+b)$. To describe a phonon wave packet of coherence length $l_c$ localized at the center of each well (e.g. $z=0$ for the central well in Fig. 1), we employ a Gaussian wave packet of half width $l_c$, described by the wave function
	\begin{equation}\label{realspacewavepacket}
		u'(z,0)=\frac{1}{q_{UV}}\int^{q_{UV}}_0 dq_z\tilde{W}'_{q_z}\frac{a}{2} e^{iq_zz} =  \frac{a}{2}e^{-\frac{z^2}{l^2_c}}
	\end{equation}
	where $\tilde{W}'_{q_z}$ is a dimensionless weight factor to be defined below, as the initial condition, chosen in such a way that the atom at $z=0$ gets displaced by half thickness of the well layer (one may consider normally smaller and more realistic amplitude, half of lattice spacing $a_{\mathrm{lattice}}$ for example; the final result for weight function is independent of this amplitude in displacement). We will work with a slightly modified weight function given by
	\[
	W_{q_z}=\frac{1}{l_c}\int^{\frac{a}{2}}_{-\frac{a}{2}}u'(z,0)e^{-iq_zz}dz=\frac{a}{2}\sqrt{\pi}\tilde{W}_{q_z}
	\]
	\begin{equation}\label{phononcoherenceweightfactor}
		=\frac{a}{4} e^{-\frac{1}{4}
			l^2_c q^2_z} \sqrt{\pi} \left(\mathrm{Erf}\left[\frac{(a - i l^2_c q_z)}{2 l_c}\right] + \mathrm{Erf}\left[\frac{(a + i l^2_c q_z)}{2 l_c}\right]\right)
	\end{equation}
	where $\mathrm{Erf}$ is error function and $\tilde{W}_{q_z}$ is defined such that $\mathrm{limit}_{l_c\rightarrow 0}\tilde{W}_{q_z}=1$. This is consistent with the constraint that incoherent phonon case corresponds to $\tilde{W}_{q_z}=1$ while giving a generally complex-valued real space function $u(z,0)$ given by Eq.(\ref{realspacewavepacket}) but with substitution $\tilde{W}'_{q_z}\rightarrow \tilde{W}_{q_z}$, as illustrated in Fig. 2A), where only the real part of $u(z,0)$ is plotted as it corresponds to physical ion displacement while the imaginary part of $u(z,0)$ does not. While the $\mathrm{Re}[u(z,0)]$ no longer looks like a Gaussian function of half width $l_c$, it still satisfies the physically intuitive picture that this function approaches more closely a Dirac delta function in real space ($\delta(z)$) as $l_c\rightarrow 0$ as its amplitude is largest in this limit. The immediate role of this coherence factor is that it leads to replacement 
	\begin{equation}
		e^{iq_zz}\rightarrow \tilde{W}_{q_z}e^{iq_zz}
	\end{equation}
	in the overlap integrals in Eq.(\ref{overlapfactor}). The phonon coherence therefore manifests simply by modifying the effective electron-phonon coupling in Eq.(\ref{Frohlich}). 
	
	Our phenomenological theory of wave packet to describe coherent phonons can also be deduced by comparison with a more rigorous microscopic theory  \cite{Lax}-\cite{LaxPRB87} formulated to construct a phonon wave function spatially localized in a well, properly adapted to define phonon coherence as conceptually proposed in \cite{VolzPRB}. The principal result of those studies \cite{Lax}-\cite{LaxPRB87} gives the phonon wave packet operator \cite{Lax} 
	\begin{equation}
		b_{\mathbf{q}_{\perp},q_z}\rightarrow \tilde{b}_{n'n}(\mathbf{q}_{\perp})\sim\sum_{q_z}G_{n'n}(\mathbf{q}_{\perp},q_z)b_{\mathbf{q}_{\perp},q_z}    
	\end{equation}
	where $n,n'=1,\cdots, N_s$ and $G_{n'n}(\mathbf{q}_{\perp},q_z)$ takes particularly simple form for $n=n'=1$
	\begin{equation}
		G_{11}(\mathbf{q}_{\perp},q_z)=\frac{1}{\sqrt{L}}\int^{\infty}_{-\infty}|\zeta_1(z)|^2dz\frac{\exp(iq_zz)}{\sqrt{q^2_{\perp}+q^2_z}}.  
	\end{equation}
	The wave packet thus indeed manifests as an effective weight factor ($\tilde{W}_q=1/\left(\sqrt{L}\sqrt{q^2_{\perp}+q^2_z}\right)$ in the context of phonon localization in the well of thickness $L$ \cite{Lax}) multiplying the factor $\exp(iq_zz)$ or, equivalently, the Frohlich electron-phonon coupling in Eq.(\ref{Frohlich}), justifying our semiclassical picture.
	
	The phonon spatial coherence enters our semiclassical theory in terms of the weight factor $\tilde{W}_{q_z}$ given in Eq.(\ref{phononcoherenceweightfactor}). Our analysis shows that having incoherent phonons gives weight factor equal unity independent of $q_z$ while having coherent phonons gives smaller weight factor, as illustrated in Fig. 2. This means coherent phonons gives weaker effective electron-phonon interaction. With this formulation, the overlap integrals in Eq.(\ref{overlapfactor}) correspond to the case with vanishing phonon coherence length $l_c\rightarrow 0$, that is, the case with completely incoherent phonons, which leads to constant weight factor $\tilde{W}_{q_z}=1$. Having coherent phonons implies large enough $l_c$(relative to the system's characteristic length scale), leading to $q_z$-dependent weight factor $\tilde{W}_{q_z}$ with profile illustrated in Fig. 3. Comparing the blue and light yellow regions suggests that having coherent LO phonons weakens electron-phonon scattering and should thus reduce carrier thermalization. This is one of the key results of our work. 
	
	\begin{figure}
		\includegraphics[angle=0,origin=c, scale=0.70]{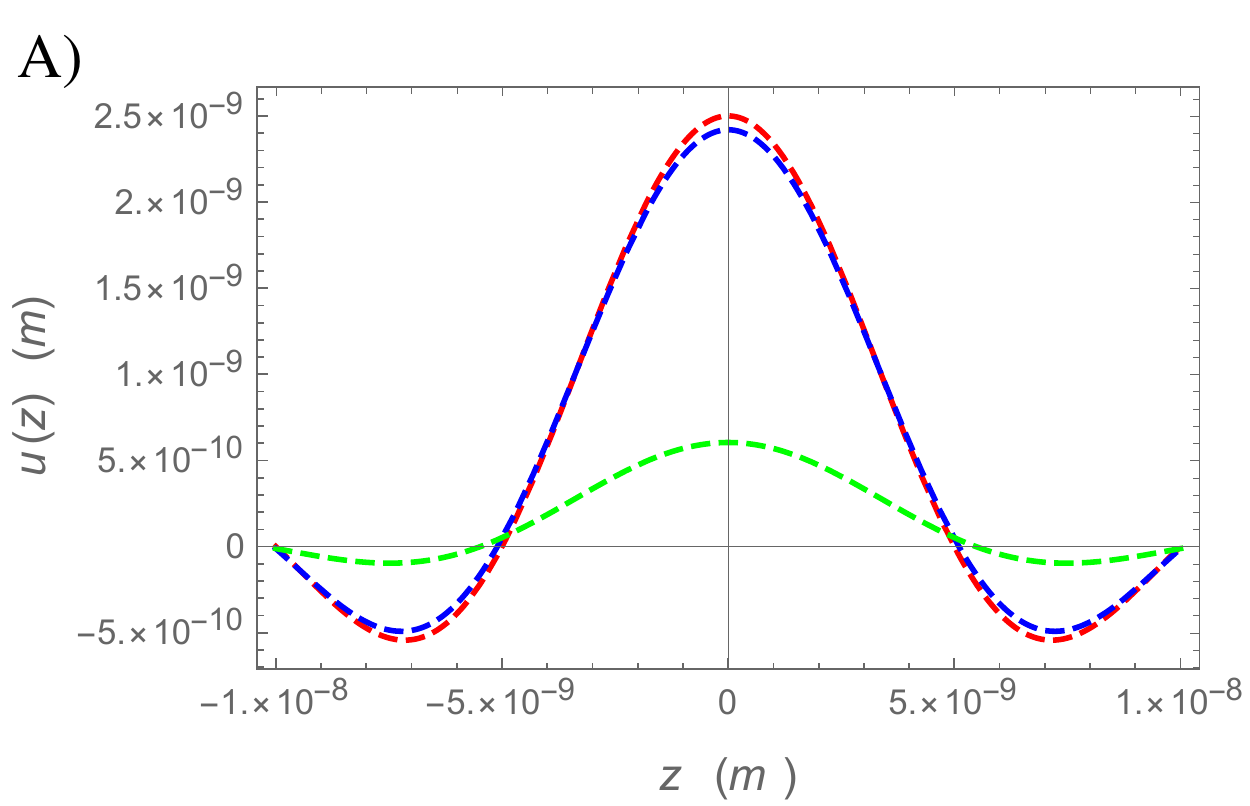}
		\includegraphics[angle=0,origin=c, scale=0.25]{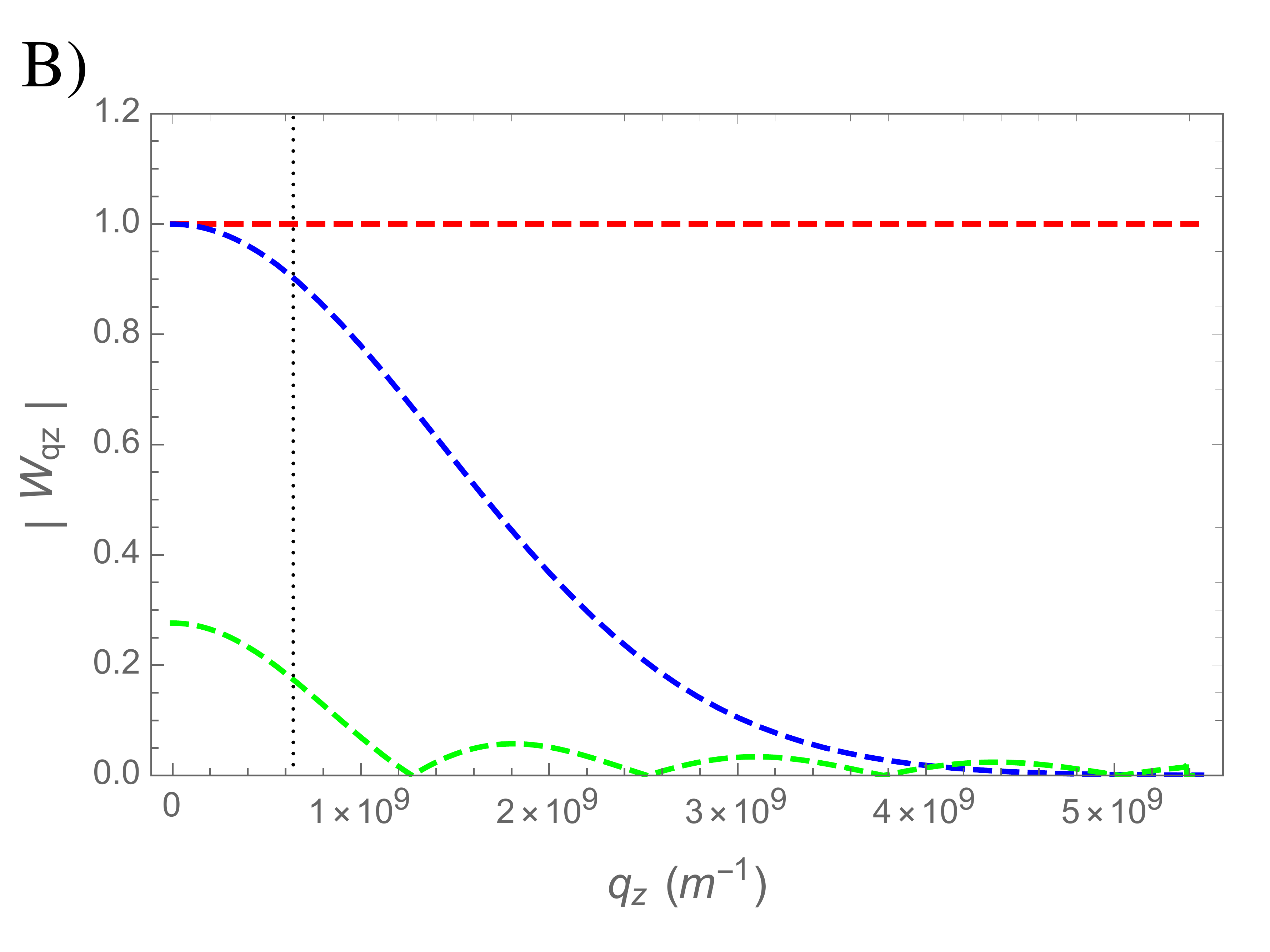}
		\label{fig:SuperlatticePhononWeightFactor}
		\caption{
			The profile of (A)the wave packet $\mathrm{Re}[u(z,0)]=(1/q_{UV})\int^{q_{UV}}_0 dq_z\tilde{W}_{q_z}(a/2)\exp(iq_zz)$ where $q_{UV}=\pi/(a+b)$ with $a=5$nm, $b=1$nm and (B) the corresponding weight factor $\tilde{W}_{q_z}$ for the case with incoherent phonon (\textcolor{red}{red})($l_c=0.000001$nm), weakly coherent (\textcolor{blue}{blue})($l_c=1$nm), and strongly coherent phonon (\textcolor{green}{green}) ($l_c= 10$nm). The vertical dashed line delimits the folded phonon Brillouin zone.}
	\end{figure}
	
	\begin{figure}
		\includegraphics[angle=-90,origin=c, scale=0.3]{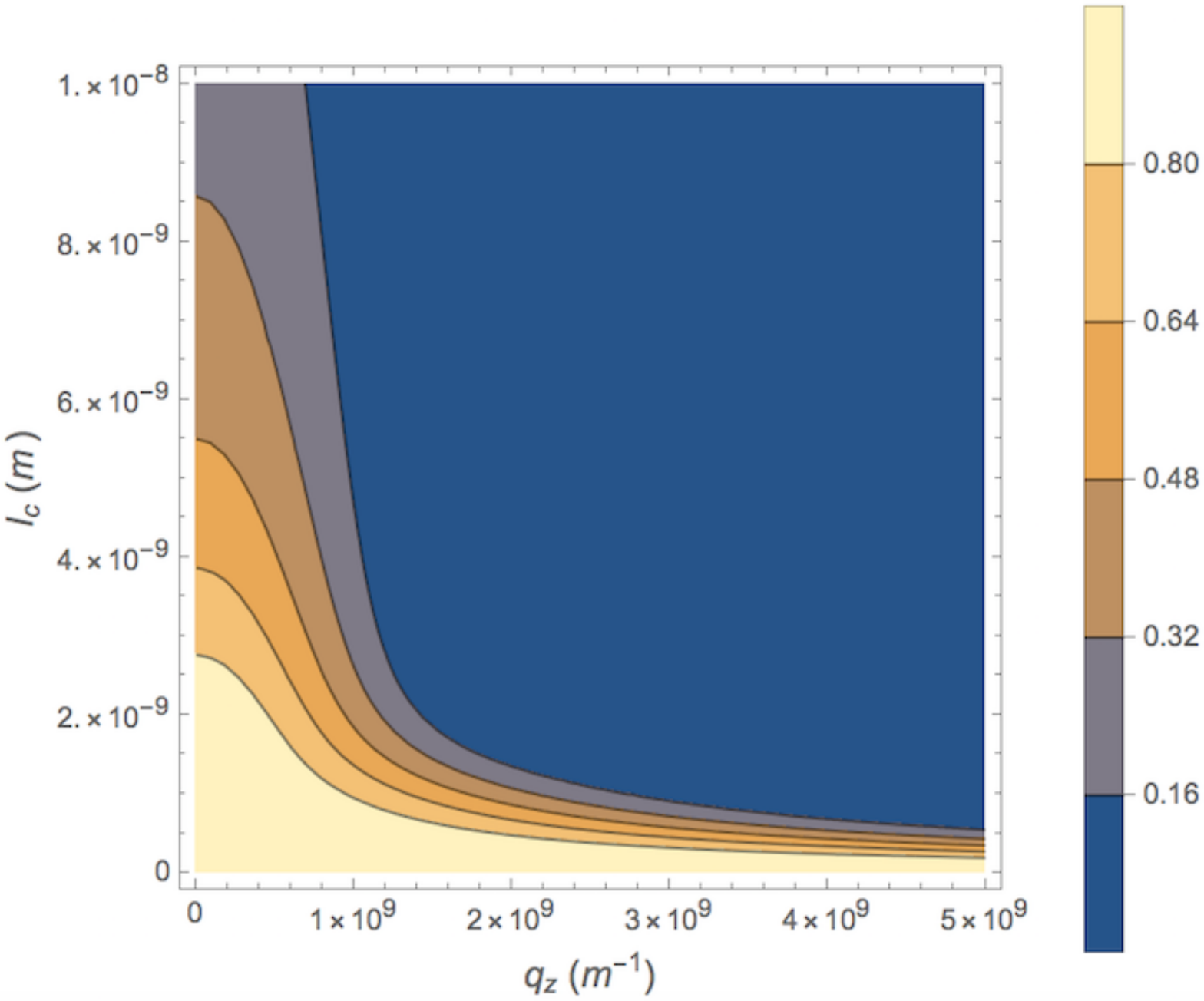}
		\label{fig:SuperlatticePhononWeightFactor}
		\caption{
			The contour plot of $|\tilde{W}_{q_z}|$ (with $a=5$nm) dissecting regions varying from coherent (blue) to incoherent (light yellow) phonons, indicating that coherent phonons give smaller effective electron-phonon coupling due to smaller $|\tilde{W}_{q_z}|$.}
	\end{figure}
	
	\section{Scattering Rate Calculation}
	The electron thermalization rate is determined in terms of thermalization power, given by the rate of energy transferred by the electrons to the longitudinal optical (LO) phonons emitted by the electrons. To that end, we compute the rate of change of the number of LO phonons, 
	\begin{equation}
		\frac{d N_{{\bf q}_\perp,q _z}}{dt} = \frac{2 \pi}{\hbar} |M_{{\bf q}_\perp,q _z} |^2 A_{\mathbf{q}_{\perp},q_z}
	\end{equation}
	where
	\[
	A_{\mathbf{q}_{\perp},q_z}=2\sum_{\mathbf{k}_{\perp}}\sum_{k_z,k'_z}I^2_{\mathbf{k},\mathbf{k}-\mathbf{q}}((N_{\mathbf{q}_{\perp},q_z}+1)f_{\mathbf{k}_{\perp},k_z}(1-f_{\mathbf{k}_{\perp}-\mathbf{q}_{\perp},k'_z})
	\]
	\begin{equation}\label{TheSumOI}
		-N_{\mathbf{q}_{\perp},q_z}f_{\mathbf{k}_{\perp}-\mathbf{q}_{\perp},k'_z}(1-f_{\mathbf{k}_{\perp},k_z}))
		\delta(\Delta E_{n\mathbf{k};n'\mathbf{k}',\mathbf{q}}-\hbar\omega_{\mathbf{q}_{\perp},q_z})
	\end{equation}
	where $\Delta E_{n\mathbf{k};n'\mathbf{k}',\mathbf{q}}=E_n(\mathbf{k}_{\perp},k_z)-E_{n'}(\mathbf{k}_{\perp}-\mathbf{q}_{\perp},k'_z)$,  ${\bf k} = ({\bf k}_\perp, k_z)$, $I^2_{\mathbf{k},\mathbf{k}-\mathbf{q}}=|G(\mathbf{k}n\mathbf{k}'n'\mathbf{q})|^2|\tilde{W}_{q_z}|^2$, $\omega_{\mathbf{q}_{\perp},q_z}=\omega_{LO}$, and $f_{\zeta}=(\exp((E_{\zeta}-\mu_c)/(k_BT_c)+1)^{-1}$; the Fermi-Dirac distribution of electron of wave vector $\zeta$ having energy $E_{\zeta}$ at temperature $T_c$ and chemical potential $\mu_c$.
	
	As shown in \cite{SupplementaryMaterials}, in the limit of large $N$ (the number of periods of the superlattice), the form factor asymptotically approaches a Dirac delta function and can be approximated as
	\begin{equation}\label{DiracDeltaFormFactor}
		|G(\mathbf{k}n\mathbf{k}'n'\mathbf{q})|^2=C_{N}(n,n',k_z,q_z)\delta(\mathbf{k}'-\mathbf{k}+\mathbf{q})
	\end{equation}
	where the function $C_N(n,n', k_z, q_z)$ is given in \cite{SupplementaryMaterials}. This Dirac delta function-type of form factor implies that the Bloch wave vector is also conserved along the transport direction of the superlattice, due to the macroscopic translational invariance arising from the superlattice periodicity. 
	Using the Dirac delta function form factor in Eq.(\ref{DiracDeltaFormFactor}), the $dN_{\mathbf{q}_{\perp},q_z}/dt$ becomes
	\[
	\frac{dN_{\mathbf{q}_{\perp},q_z}}{dt}
	=\frac{N_{\mathbf{q}}(T_c) - N_{\mathbf{q}}}{\tau^{c-LO}_{\mathbf{q}}}=(N_{\mathbf{q}}(T_c) - N_{\mathbf{q}})
	\]
	\[\times\left[2N\frac{1}{2\pi\hbar}\left(\frac{e^2\hbar\omega_{\mathbf{q}}}{q_{\perp} }\left[\frac{1}{\varepsilon_{\infty}}-\frac{1}{\varepsilon_s}\right]\frac{1}{8\pi q^2 V}\right)\right]\frac{2m_c}{\hbar^2}\frac{S}{(2\pi)^2}
	\]
	\[  
	\times(|\tilde{W}_{q_z}|)^2(a+b)^2\sum_{n,n'}   \int^{\frac{\pi}{(a+b)}+q_z}_{-\frac{\pi}{(a+b)}+q_z}dk_zC_N(n,n',k_z, q_z)
	\]
	\begin{equation}\label{rateofchangeofbosonnumber}
		\times \frac 1 2 \frac{\sqrt{2 m_c k_B T_c}}{\hbar}   \Gamma \Bigl( \frac 1 2\Bigr ) (Li_{1/2}(- e^{x_2}) -Li_{1/2}(-e^{x_1}) )
	\end{equation}
	where $N_{\mathbf{q}}(T_c)=(\exp(\hbar\omega_{\mathbf{q}}/(k_BT_c))-1)^{-1}$, $\omega_{\mathbf{q}}=\omega_{LO}$, $T_c$ is the carrier (electron) temperature, $S$ is the transverse cross sectional area of the superlattice, $\Gamma(x)$ is gamma function, and $Li_{\alpha}(x)$ is polylogarithmic function of order $\alpha$ while $x_1$ and $x_2$ are given by
	\[
	x_1 =-\beta \Bigl[\frac{\hbar^2 }{2 m_c}(k^{min}_\perp)^2+E_{n'}(a) -2V_{n'}\cos (k_z-q_z)(a+b)\Big]
	\]
	\begin{equation}\label{x1}
		+\beta (\mu_c+\hbar \omega_{LO}),
	\end{equation}
	\begin{equation}\label{x2}
		x_2= -\beta \Bigl[\frac{\hbar^2 }{2 m_c}(k^{min}_\perp)^2+E_n(a) -2V_{n}\cos k_z(a+b) -\mu_c\Big]
	\end{equation}
	where $k^{min}_{\perp}$ is the minimum transverse wave vector an electron must posses to be able to emit an LO phonon, 
	\[
	k^{\mathrm{min}}_{\perp}(n,n')= \frac{q_\perp}{2 } +\frac{(E_{n'}(a)-E_n(a))}{2 q_\perp}\frac{2m_c}{\hbar^2} 
	+ \frac{m_c \omega_{q_{\perp},q_z} }{\hbar  q_\perp}
	\]
	\begin{equation}
		- \frac{m_c}{\hbar^2q_{\perp}}\left(2V_{n'}\cos k'_z(a+b)-2V_{n}\cos k_z(a+b)\right)
	\end{equation}
	and $\mu_c=\mu_e$ the electron chemical potential. One can see that the scattering rate is proportional to $N$; the number of periods in the superlattice. Only the scattering rate per period (consisting of one well and one barrier) will be concerned. 
	
	\section{Thermalization Power}The scattering rate is used to compute the thermalization power, defined as the power needed to keep the electrons at a given effective temperature $T_c$. In steady state, this quantity equals the rate of energy lost by the electrons to LO phonons, computed from $dN_{\mathbf{q}_{\perp},q_z}/dt$ in Eqs.(\ref{rateofchangeofbosonnumber}). The thermalization power (per well) is given by 
	\begin{equation}\label{thermalpower3dbulk}
		P^{}_{\mathrm{th}}=a\int \frac{d^3q}{(2\pi)^3} \hbar\omega_{\mathbf{q}} \frac{N_{\mathbf{q}}(T_c)-N_{\mathbf{q}}(T_L)}{\tau^{c-LO}_{\mathbf{q}}+\tau^{LO-ac}_{\mathbf{q}}}
	\end{equation}
	in units of Watt/cm$^2$ describing the rate of energy loss of the electrons from emitting LO phonons within the well layer. The form of thermalization power in Eq.(\ref{thermalpower3dbulk}) reflects non-equilibrium hot LO phonon effect in steady-state \cite{Tsai}.
	
	The numerical results for electron thermalization power $P^{}_{\mathrm{th}}$ and its dependencies on well layer and barrier layer thicknesses are presented in Fig. 4, for incoherent and coherent phonons. Numerical simulation demonstrates that $l_c$ for optical modes is dependent on the superlattice period $d_{SL}=a+b$, with ``leading order" dependence of the form $l_c=\xi d_{SL}$ where $\xi$ is a (numerically-computed) function of energy of the mode and $d_{SL}$ itself  \cite{LatourChalopinPRB}; we take $\xi=2$ for simple illustration (as we fix the energy to $E_0$, the actual $\alpha$ should in principle still have some dependence on $d_{SL}$). More generally, coherent phonons correspond to $\xi\geq 1$ \cite{LatourChalopinPRB} and our results apply as long as this constraint is satisfied. Fig. 4A) demonstrates that in a superlattice heterostructure, the thermalization power decreases with the barrier thickness increase. This is because with thicker barrier, the electron is more confined, and stronger confinement reduces carrier thermalization, as is the case as well with a quantum well \cite{PRapplied}. The hot carrier effect is thus enhanced in a superlattice structure as one increases the barrier thickness because the power needed to sustain a given electron effective temperature is lower, at larger barrier thickness.
	
	\begin{figure}
		\includegraphics[angle=0,origin=c, scale=0.28]{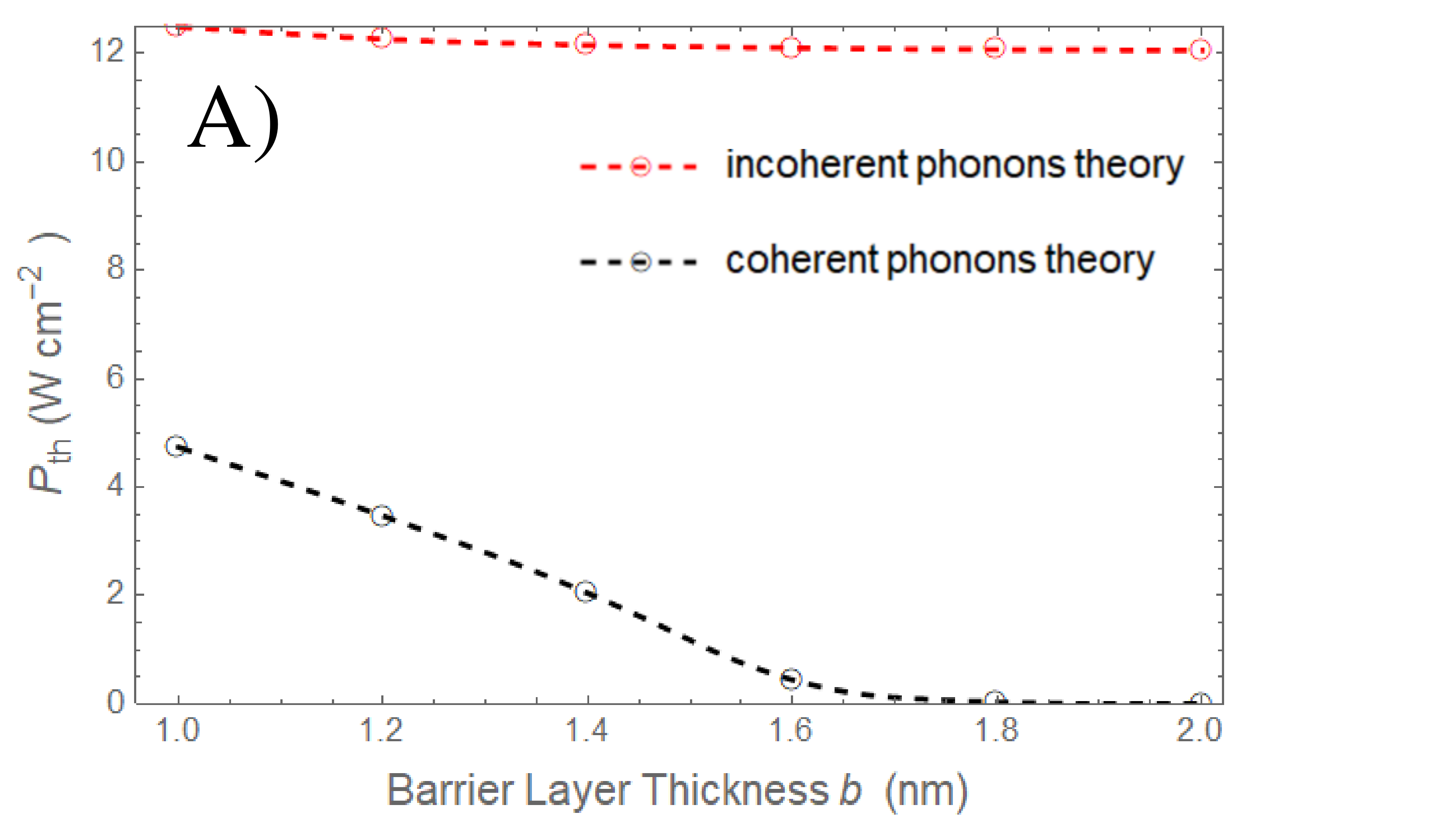}
		\includegraphics[angle=0,origin=c, scale=0.28]{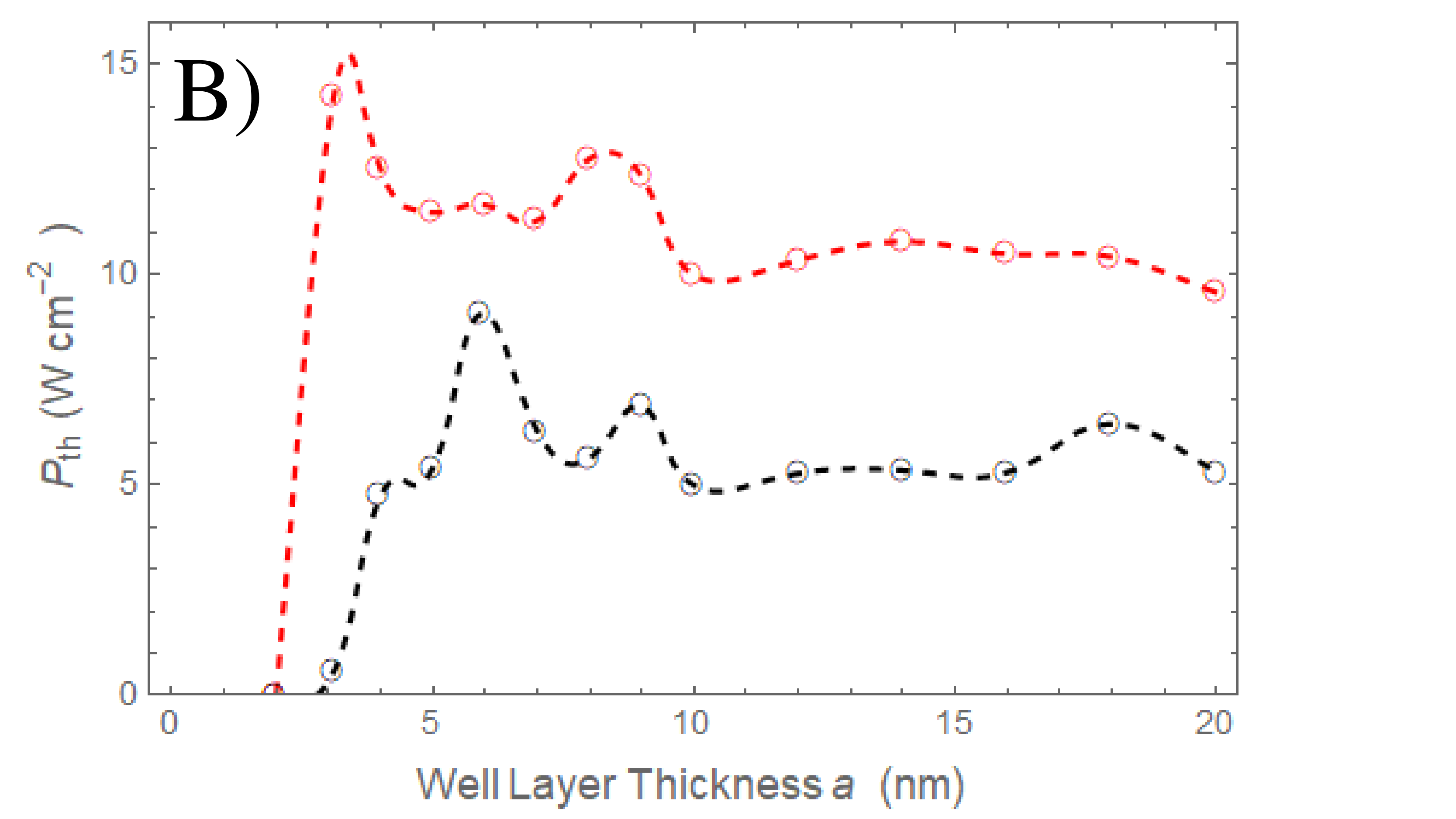}
		\label{fig:ThermalizationPowervsBarrierWellThicknesses}
			\begin{center}
				\caption{
					The theoretical dependencies of the thermalization power (in W/cm$^2$) per well of the superlattice: (A) $P_{\mathrm{th}}$ as a function of barrier thickness $b$ at fixed well layer thickness $a=4$ nm (chosen to give two mini-subbands for an InAs/AlAsSb superlattice) in terms of $P_{\mathrm{th}}$ (B) $P_{\mathrm{th}}$ as a function of well thickness $a$ at fixed barrier thickness $b=1$ nm, for LO phonon coherence length $l_{c}=10^{-6}$nm (\textcolor{red}{red})(incoherent phonons) and $l_{c}=2(a+b)$nm (\textcolor{black}{black})(coherent phonons). All other parameter values are given in \cite{SupplementaryMaterials}. The dashed lines serving as guides to the eye are obtained from interpolations of data points marked by the circles.}
			\end{center}
		\end{figure}
		
		Fig. 4A) also indicates that coherent phonons lower the thermalization power (by as much as $(P_{\mathrm{th}}(\mathrm{incoh})-P_{\mathrm{th}}(\mathrm{coh}))/P_{\mathrm{th}}(\mathrm{incoh})\simeq 62.8\%$ for 4 nm well layer thickness and barrier layer thickness of 1 nm in an InAs/AlAsSb superlattice) and thus strongly enhances the hot carrier effect. Fig. 4B) also demonstrates that thermalization power is strongly suppressed (hot carrier effect is significantly enhanced) by coherent phonons, especially for thin enough well layer. A critical well layer thickness can be defined (about 4 nm in our example of InAs/AlAsSb superlattice, corresponding to 2 mini-subbands); for thinner wells, coherent phonons can strongly reduce the thermalization power (by one order of magnitude or more for well layer thickness of 3 nm or smaller), while for thicker wells, phonon coherence effect is not as strong. This strong enhancement of hot carrier effect at small well layer thickness $a$(and thus small $d_{SL}$ for fixed $b$) is furthermore enforced by the fact that the actual numerically-determined $\alpha$ turns out to be larger at smaller $d_{SL}$\cite{LatourChalopinPRB}; the black curve would therefore be even more lowered for small $a$ while raised for large $a$. The hot carrier effect in superlattice solar cell is thus maximally enhanced when electrons are strongly confined and the LO phonons are coherent at the same time. This is another key result of our paper. 
		
		\begin{figure}
			\includegraphics[angle=0,origin=c, scale=0.3]{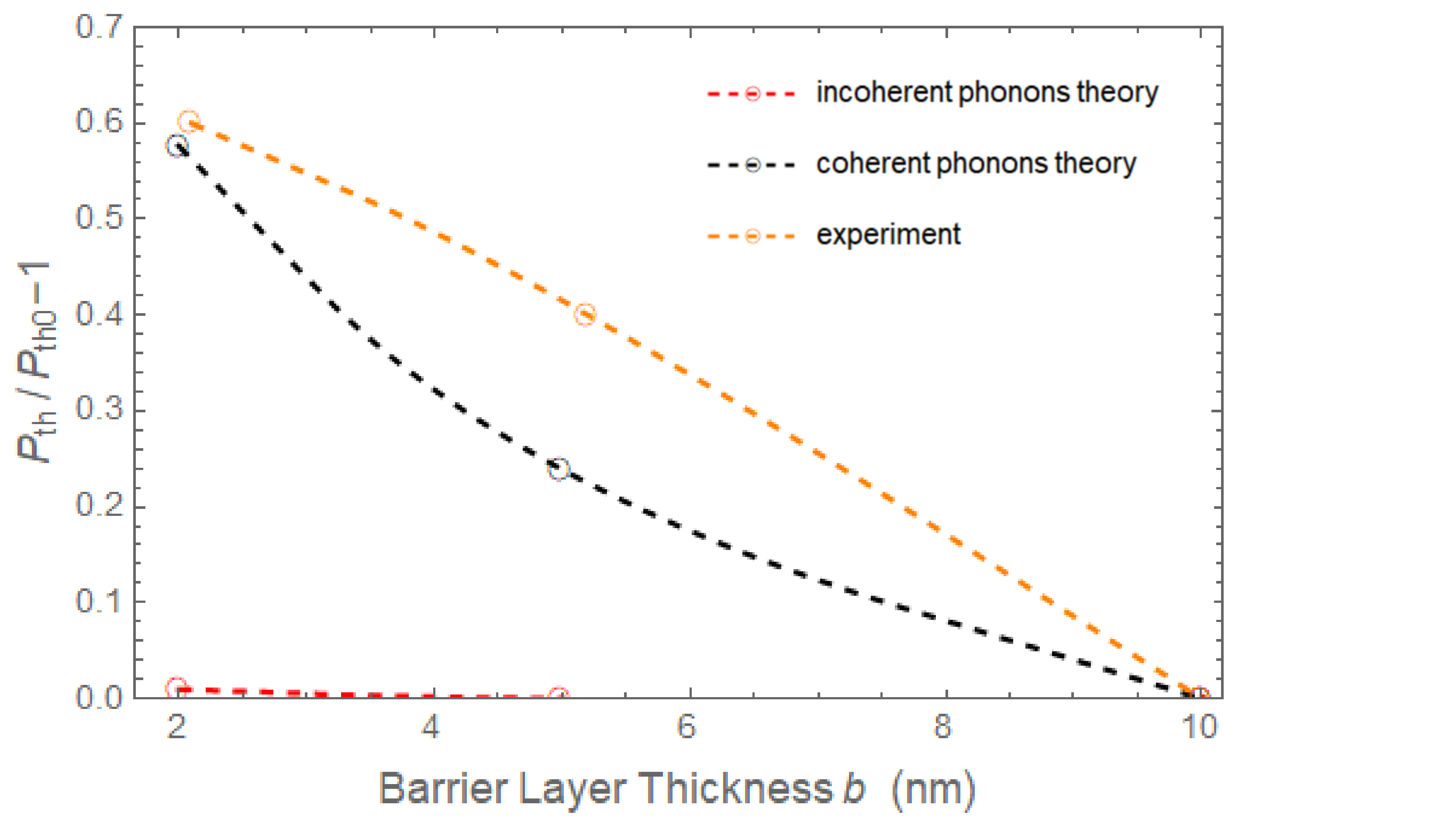}
			\label{fig:ThermalPowerTheoryvsExperiment}
			\caption{
				The normalized thermalization power dependence on barrier thickness from our theory with coherent phonons of $l_c=2d_{SL}=2(a+b)$ (black), incoherent phonons ($l_c=10^{-6}$nm)(\textcolor{red}{red}) and experiment (\textcolor{orange}{orange})\cite{EsmaielpourAPL}. Here, $P_{\mathrm{th0}}=P_{\mathrm{th}}(b\gg a)$.}
		\end{figure}
		
		\section{Comparison with Experiment}
		In this section, we provide evidence that the coherent phonons-driven hot carrier effect predicted in the present work has in fact been observed in an experiment on a superlattice solar cell prototype studied in \cite{EsmaielpourAPL}.
		
		Our theoretical result showing that the thermalization power at fixed well layer thickness decreases with the increase of barrier thickness, as shown in Fig. 4A), agrees with experimental observation \cite{EsmaielpourAPL}. However, the amount of the decrease of $P_{\mathrm{th}}$ with $b$ at fixed $a$ for the case with incoherent phonon, as shown by the red curve in Fig. 4A), while in accord with expectation, is very weak, being at least one order of magnitude smaller than its ``principal value", and also weaker than that observed in experiment \cite{EsmaielpourAPL}. Noting from Eq.(\ref{phononcoherenceweightfactor}) that the phonon coherence-induced weight factor $\tilde{W}_{q_z}$ has no explicit dependence on barrier thickness $b$, this simply implies that electron confinement effect alone cannot account for the decrease in thermalization power with barrier thickness and confirms that the phonon coherence length $l_c$ must indeed depend on the barrier thickness. Our result thus verifies existing theoretical works that show that the phonon coherence length $l_c$ in a superlattice is dependent on its period $d_{SL}=a+b$ \cite{VolzPRB}\cite{LatourChalopinPRB}; more precisely, for optical mode, $l_c$ is dependent on the superlattice period $l_c\gtrsim d_{SL} =(a+b)$ \cite{LatourChalopinPRB}. 
		
		To make a more quantitative comparison with the experiment, the thermalization power $P_{\mathrm{th}}$ is computed at a fixed well layer thickness $a=2$nm and three barrier layer thicknesses $b$'s used in \cite{EsmaielpourAPL}. Just like in previous section, $l_c=2(a+b)$ is used for illustration, corresponding to $\xi=2$, indicating rather strongly coherent phonons. A quantity called ``normalized thermalization power" is defined, given by $P_{\mathrm{th}}/P_{\mathrm{th0}}-1$, where $P_{\mathrm{th0}}=P_{\mathrm{th}}(b\gg a)$, to compare theory and experiment. The reason of doing this is at least two folds; first our theory is derived for infinitely long superlattice $N\rightarrow\infty$ while available experiments clearly studied finite-length superlattice ($N$ is finite). Second, our theory only considers the so-called ```volumic" contribution to thermalization power whereas a full description of experimental system should cover both volumic and surfacial contributions. Therefore, we cannot compare the absolute values of thermalization between theory and experiment. Instead, the normalized thermalization power as defined earlier better captures the dependence of thermalization power on barrier layer thickness, excluding the surface contribution, which is normally additive \cite{GiteauJAP} and minimizing finite-size($N$) effect.  
		
		As demonstrated in Fig. 5, the decrease in thermalization power with barrier thickness observed in \cite{EsmaielpourAPL} cannot be reproduced (in the strength of its dependence on barrier thickness) by assuming incoherent phonons, while assuming coherent phonons (with $l_c=2(a+b)$ for illustration) does reflect the experimental data reasonably well, with residual discrepancy that may originate from simplifications involved our theory. In other words, our theory suggests that the hot carrier effect observed in the superlattice heterostructure for solar cell in \cite{EsmaielpourAPL} is largely driven by coherent phonons, which are activated by the laser irradiation.  Altogether, this result indicates that, while both electron confinement and phonon coherence leads to reduction of thermalization power, and thus enhancement of hot carrier effect, the contribution of phonon coherence can in fact be significant or even dominant, under steady-state irradiation situation, as implemented in \cite{EsmaielpourAPL} and as is also the standard condition in solar cell operation.  
		
		\section{Discussion and Conclusions}
		Our work presents a minimal model to describe carrier thermalization in a superlattice solar cell with electron-phonon scattering as the main channel of electron energy loss. The role of an inherently quantum mechanical property of phonons; the quantum coherence, is unraveled, and its impact on the carrier thermalization, a process that crucially determines the solar cell efficiency, is investigated. Our results provide a proof of principle that having coherent phonons would enhance hot carrier effect in superlattice solar cell and, prospectively, its efficiency, complementing also the use of photon quantum coherence \cite{ScullyPRL} to achieve the same objective. From the phonon's perspective, our finding that longer phonon coherence length gives lower thermalization power implies longer phonon life time, a conclusion that fully agrees with existing works \cite{GargChen}\cite{AdvancesinPhysics}. Our analysis suggests that the hot carrier effect observed in \cite{EsmaielpourAPL} is largely driven by coherent phonons, indicating that phonon coherence can have dominant contribution in suppressing carrier thermalization under steady-state irradiation as in standard solar cell operation. 
		
		Our calculations implicitly assumed that the LO phonons are confined (or localized) in the wells, periodically arranged along the superlattice. This is justified by the fact that the difference in the polarizabilities of the semiconductor materials that constitute the well and barrier layers acts as phononic confining potential for the optical phonons. While this assumption simplifies the calculations, the critical final results of coherent phonons-driven hot carrier effect does not depend on this confined LO phonons assumption. In fact, an alternative scenario with extended LO phonons picture would still preserve the coherent phonons-driven hot effect, as can be deduced by making comparison to the study on a quantum well solar cell \cite{PRapplied}. The acoustic phonons, which are indeed not polar and therefore not sensitive to the difference in polarizabilities, remain extended, that is, not localized. Since optical phonons have vanishing group velocity and thus do not carry heat while acoustic phonons travel at sound velocity (at small wave vector or, equivalently, long distance regime) and thus capable of carrying heat over a long distance, this means the superlattice still maintains high thermal conductivity even within our localized optical phonons picture. In other words, our theory implies that the electrons (and holes) remain energetic (hot) because they do not lose much energy as their scatterings with optical phonons are reduced by the coherence of the latter, without reducing the overall thermal conductivity of the superlattice, thus preserving the capacity to remove excess heat. Indeed, any lost kinetic energy of the electrons due to the scattering with optical phonons and any excess phononic thermal energy due to temperature gradient can still be transferred via phonon-phonon scattering to acoustic phonons, who then efficiently transport the heat across the superlattice, because the acoustic phonons remain extended. Our results thus offer a new approach to minimize carrier energy loss while at the same time maintaining high thermal conductivity for effective excess heat removal process; two qualities that are desirable for high efficiency solar cell systems.
		
		Ab-initio calculation \cite{LatourChalopinPRB} suggests that the coherence length $l_c$ depends on the superlattice period and energy of the phonon mode. As such, $l_c$ is only weakly or indirectly dependent on the material choice, through the dependence of the phonon energy of the material parameters. Our work pertains principally to polar semiconductor materials because in such materials, the optical phonon plays the direct role in the electron-phonon scattering that is responsible for the carrier thermalization. Now, \cite{LatourChalopinPRB} shows that $l_c/d_{SL}$ in general decreases with energy. This means coherence length $l_c$ for the optical phonon is maximized if the optical phonon energy is minimized. This provides a guideline for the material choice. The energy or frequency of the optical phonon mode must be minimized. This condition can be investigated using the linear diatomic chain model of phonon. In such model, suitable to describe diatomic III-V or II-VI semiconductors such as GaAs, InP, ... the optical phonon energy (at $q_z=0$) is given by $E_0 = 2 \hbar K (1/m_1 + 1/m_2)$ where $K$ is an effective spring constant of the force between the ions while $m_1$ and $m_2$ are atomic masses of the two ions of the elements constituting the compound semiconductor. While the effective spring constant $K$ may have delicate dependence on the atomic configuration and potential energy, the dependence on the masses clearly suggests that the optical phonon energy is minimized when using elements of largest ion masses $m_1$ and $m_2$. For III-V family, this would be InSb compound, while for II-VI family, this would be SrTe (excluding the heavier elements in periodic table in these two groups which are not normally used for semiconductor materials as they are radioactive). Certainly, these choices of materials must be consolidated with the electronic band gap requirement needed to form superlattice and the final choice of materials may involve trade off or compromise between these different requirements and can also necessitate the use of alloys of different compound semiconductors. 
		
		As a final note, while our theory is formulated in terms of superlattice due to the resulting analytical simplicity, the conclusions are expected to apply as well to multi-quantum wells heterostructure, as the former can be taken as the extension of the latter to infinitely long and perfectly periodic structure; both structures are extensively used for solar cell design. Our present work should therefore motivate further theoretical and experimental studies in cultivating quantum mechanical properties of phonons to improve the performance of photovoltaic devices.
		
		\section{Acknowledgements} 
		I.M. acknowledges funding from the French Agence Nationale de la Recherche (ANR) through project ANR-ICEMAN under Grant Number:19-CE05-0019-01 during his work at IM2NP Marseille. This project has received funding from the European Union’s Horizon 2020 research and innovation programme under the Marie Skłodowska-Curie grant agreement No 899987 (H.E.).

		\begin{widetext}
			
			\centering
			\textbf{Supplementary Materials: Coherent Phonons-Driven Hot Carrier Effect in Superlattice Solar Cell}\\
			\centering
			I. Makhfudz$^{1}$, N. Cavassilas$^{1}$, Y. Hajati$^{2}$, H. Esmaielpour$^3$, and F. Michelini$^{1}$
			
				\centering
			$^{1}$IM2NP, UMR CNRS 7334, Aix-Marseille Universit\'{e}, 13013 Marseille, France\\
		$^{2}$Department of Physics, Faculty of Science, Shahid Chamran University of Ahvaz, 6135743135 Ahvaz, Iran\\
		$^3$Walter Schottky Institut, Technische Universität München, Am Coulombwall 4, D-85748 Garching, Germany\\
		\comment{$^{3}$Institut für Physik, Martin-Luther-Universität Halle-Wittenberg, 06099 Halle (Saale), Germany}
		\date{\today}
		\bigskip
		
		In these Supplementary Materials, we provide additional details supplementing the main text. First, we present a derivation of the electron density of states, useful for calculation of electron density given its chemical potential. Next, we give the details of the form factor appearing in the electron-phonon Hamiltonian. Third, we describe the derivation of the electron-phonon scattering rate. The file ends with details on material and operational parameters used in the calculations.
		
	\end{widetext}
	
	\section{Electron Density of States in Superlattice}
	
	\begin{figure}
		\includegraphics[angle=0,origin=c, scale=0.5]{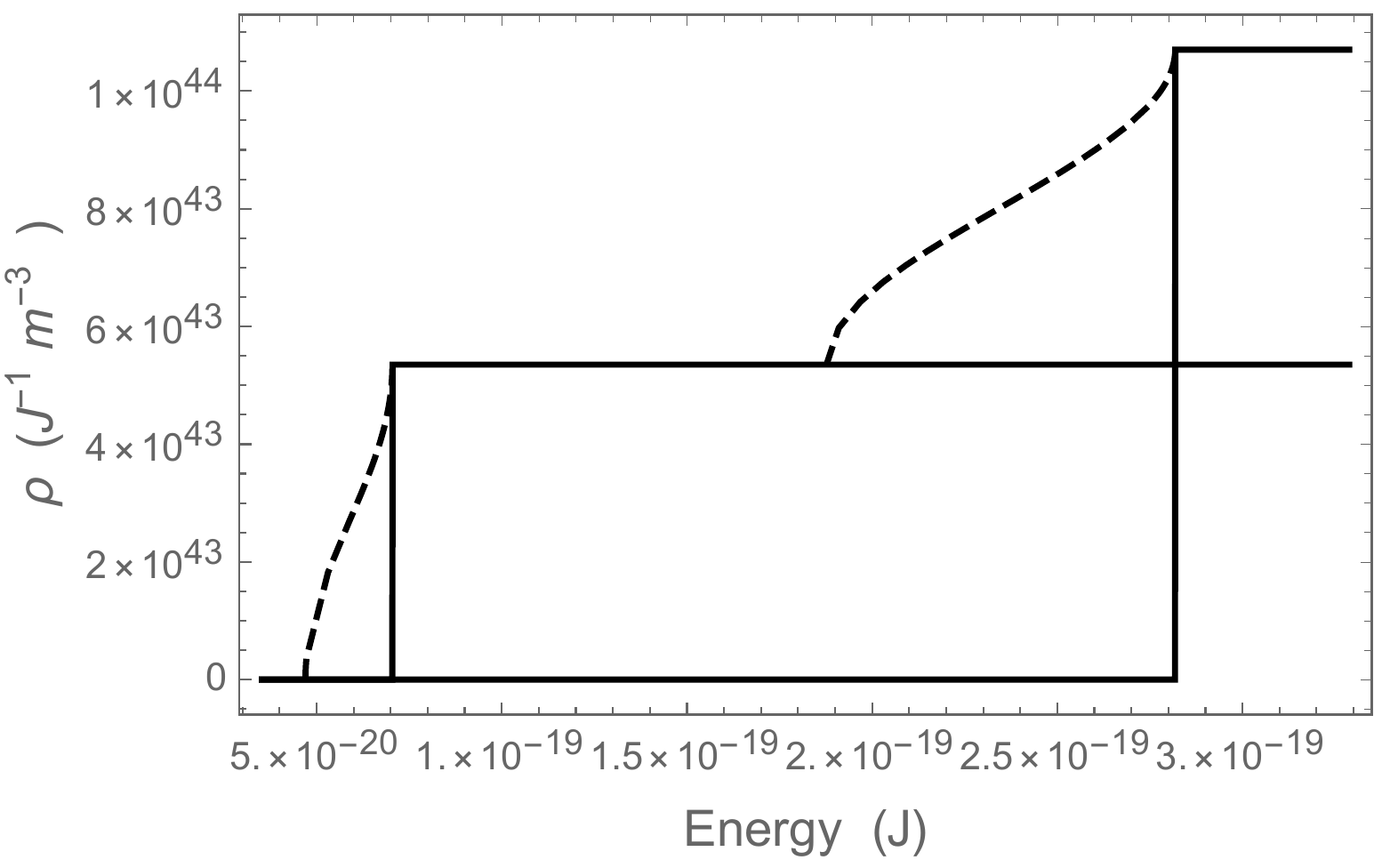}
		\label{fig:ElectronDensityofStates}
		\caption{
			The profile of the electron density of states in a superlattice consisting of wells of width $a$ and barriers of width $b$. The dashed curves represent the mini-subbands.}
	\end{figure}
	
	From the energy dispersion Eq.(3) in the main text, one can compute the density of states given by (not including the spin degree of freedom)
	\begin{equation}\label{DOSdefinition}
		\rho(E)=\sum^{N_s}_{n=1}\int \frac{d^3k}{(2\pi)^3}\delta(E-E_n(\mathbf{k}))
	\end{equation}
	where $N_s$ is the number of mini-subbands. Using the energy dispersion
	\begin{equation}\label{superlatticedispersionS}
		E_n(\mathbf{k})=E_n(k_{\perp})-2V_n\cos k_z (a+b)
	\end{equation}
	tedious but otherwise straightforward evaluation of the integral in Eq.(\ref{DOSdefinition}) gives
	\begin{widetext}
		\begin{equation}\label{DOSfunction}
			\rho(E) =\frac{dN(E)}{dE}=
			\begin{cases}
				\frac{m_c}{2\pi^2\hbar^2}\sum_n\left(k^{\mathrm{max}}_z(n, E)-k^{\mathrm{min}}_z(n, E)\right) & \text{if $0<k^{\mathrm{max}}_z(n, E),k^{\mathrm{min}}_z(n, E)<\frac{\pi}{(a+b)}$}\\
				\frac{m_c}{2\pi^2\hbar^2}\sum_nk^{\mathrm{max}}_z(n, E) & \text{if $k^{\mathrm{min}}_z(n, E)<0<k^{\mathrm{max}}_z(n, E)<\frac{\pi}{(a+b)}$}
				\\
				\frac{m_c}{2\pi^2\hbar^2}\sum_n\frac{\pi}{(a+b)} & \text{if $k^{\mathrm{min}}_z(n, E)<0<\frac{\pi}{(a+b)}<k^{\mathrm{max}}_z(n, E)$}
			\end{cases}       
		\end{equation}
	\end{widetext}
	where
	\begin{equation}
		k^{\mathrm{min}}_z(n, E)=\frac{1}{(a+b)}\arccos\left[\frac{\hbar^2}{4m_cV_n}\left(\frac{n\pi}{a}\right)^2-\frac{E}{2V_n}\right],
	\end{equation}
	\begin{equation}
		k^{\mathrm{max}}_z(n, E)=\frac{1}{(a+b)}\arccos\left[\frac{\hbar^2\left({k^{\mathrm{max}}_{\perp}}^2+\left(\frac{n\pi}{a}\right)^2\right)}{4m_cV_n}-\frac{E}{2V_n}\right]
	\end{equation}
	noting that $\rho(E)$ involves a sum over densities of states of all mini-subbands. In the above expression, $k^{\mathrm{max}}_{\perp}$ is the upper (ultraviolet UV) cutoff on the transverse wave vector of the electron (that is, the 2D wave vector defined in the plane of the layer of the heterostructure). We will take 
	\begin{equation}
		k^{\mathrm{max}}_{\perp}=\Lambda_{\perp}=\frac{\pi}{a_{\mathrm{lattice}}}
	\end{equation}
	where $a_{\mathrm{lattice}}$ is the lattice spacing of the crystal of the semiconductor quantum well layer.
	
	The first case in Eq.(\ref{DOSfunction})  is equivalent to zero density of states because $\left(k^{\mathrm{max}}_z(n)-k^{\mathrm{min}}_z(n)\right)<0$ while a density of states can never be negative. The remaining (second and third) possibilities give a compact expression for the density of states as
	\begin{widetext}
		\begin{equation}
			\rho(E)=  \frac{m_c}{2\pi^2\hbar^2}\sum_n\left(\frac{\pi}{(a+b)}\Theta(E-(E_n(\Lambda_{\perp})+2V_n))+k^{\mathrm{max}}_z(n,E)(\Theta(E-(E_n(\Lambda_{\perp})-2V_n))-\Theta(E-(E_n(\Lambda_{\perp})+2V_n)))\right)
		\end{equation}
	\end{widetext}
	with a profile shown in Fig. above. The density of states of holes is obtained in similar manner. These densities of states are then used in the calculation of chemical potentials $\mu_e (\mu_h)$ for electron (hole) respectively, related by quasi-Fermi level splitting $\Delta\mu = \mu_e-\mu_h$, using the charge neutrality condition $n_d=p_d$, where $n_d, p_d$ are respectively the electron and (heavy) hole densities \cite{MakhfudzJPDAP}, that describes the intrinsic semiconductor (InAs in our example) that acts as the well layer. The details of the computation of chemical potential from densities of states and charge neutrality condition follow closely that presented in \cite{MakhfudzJPDAP} and will not be repeated here.
	
	\section{Evaluation of Overlap Integral Form Factor}
	Evaluating the overlap integral form factor in Eq.(7) in the main text gives
	\[
	G(\mathbf{k}n\mathbf{k}'n'\mathbf{q})   =\frac{\delta(\mathbf{k}'_{\perp}-\mathbf{k}_{\perp}+\mathbf{q}_{\perp})}{N}
	\]
	\begin{equation}\label{overlapfactorS}
		\times \sum^N_{\alpha,\alpha'=1}e^{i(k'_zz_{\alpha'}-k_zz_{\alpha})}G(n\alpha n'\alpha' q_z)
	\end{equation}
	and 
	\begin{equation}\label{overlapintegral}
		G(n\alpha n'\alpha' q_z)=\int dz \phi^*_n(z-z_{\alpha})e^{iq_zz}\phi_{n'}(z-z_{\alpha'})
	\end{equation}
	where $z_{\alpha}=\alpha(a+b),z'_{\alpha}=\alpha'(a+b)$. As stated in Eq.(8) in the main text, we have employed analytical wave function; the eigenfunction of an infinite potential well, to compute the overlap integral form factor, as it permits exposition of analytical properties of electron-phonon interaction in superlattice problem, especially those emerging from the spatial periodicity in a superlattice. This approximation is good enough for relatively deep well (or tall barrier), as is the case for the InAs/AlAsSb studied in present work. While the eigenfunction Eq.(8) in the main text has zero overlap between two adjacent wells, the effect of wave function overlap in the barrier region is contained in the tunneling energy $V_n$. Within this analytical approximate formulation, only $\alpha=\alpha'$ contribute to the sum in Eq.(\ref{overlapfactorS}). In this case, the interval of integration over $z$ is $z\in (\alpha(a+b)-a/2,\alpha(a+b)+a/2)$. Within our model, $V=Sa$ where $S$ is the transverse cross sectional area of the superlattice. 
	
	The Dirac delta in Eq.(\ref{overlapfactorS}) imposes conservation of transverse wave vector $\mathbf{k}'_{\perp}=\mathbf{k}_{\perp}-\mathbf{q}_{\perp}$. Evaluating the integral, we obtain
	\[
	G(\mathbf{k}n\mathbf{k}'n'\mathbf{q}) =
	\frac{\delta(\mathbf{k}'_{\perp}-\mathbf{k}_{\perp}+\mathbf{q}_{\perp})}{N}\mathcal{G}(k_zn k'_zn'q_z)
	\]
	where $\mathcal{G}(k_zn k'_zn'q_z)$ is a dimensionless complex function 
	\begin{widetext}
		\[
		\mathcal{G}(k_zn k'_zn'q_z)=
		\sum^N
		_{\alpha,\alpha'=1}e^{i(k'_z\alpha'-k_z\alpha)(a+b)}
		\left(-e^{
			i (a (-(\frac{1}{2}) + \alpha) + \alpha b) q_z} \left(\frac{(
			i a q_z \cos[\frac{1}{2} (n - n') \pi] + (-n + n') \pi \sin[
			\frac{1}{2} (n - n') \pi])}{((n \pi - n' \pi - a q_z) (n \pi - 
			n' \pi + a q_z))}\right) \right)
		\]
		\[
		+        \sum^N
		_{\alpha,\alpha'=1}e^{i(k'_z\alpha'-k_z\alpha)(a+b)}
		\left(-e^{
			i (a (-(\frac{1}{2}) + \alpha) + \alpha b) q_z} \left(\frac{(
			i a q_z \cos[\frac{1}{2} (n + n') \pi] - (n + n') \pi \sin[
			\frac{1}{2} (n + n') \pi])}{((n \pi + n' \pi - a q_z) (n \pi + 
			n' \pi + a q_z))}\right) \right)
		\]
		\[
		+\sum^N
		_{\alpha,\alpha'=1}e^{i(k'_z\alpha'-k_z\alpha)(a+b)}
		\left(
		e^{i (a (\frac{1}{2} + \alpha) + \alpha b) q_z} \left(\frac{(
			i a q_z \cos[\frac{1}{2} (n - n') \pi] + (n - n') \pi \sin[
			\frac{1}{2} (n - n') \pi])}{((n \pi - n' \pi - a q_z) (n \pi - 
			n' \pi + a q_z))} \right)\right)
		\]
		\begin{equation}
			+\sum^N
			_{\alpha,\alpha'=1}e^{i(k'_z\alpha'-k_z\alpha)(a+b)}
			\left(
			e^{i (a (\frac{1}{2} + \alpha) + \alpha b) q_z} \left(\frac{(
				i a q_z \cos[\frac{1}{2} (n + n') \pi] + (n + n') \pi \sin[
				\frac{1}{2} (n + n') \pi])}{((n \pi + n' \pi - a q_z) (n \pi + 
				n' \pi + a q_z))}\right)\right).
		\end{equation}
	\end{widetext}

	
	Evaluating the sum over $\alpha$, we obtain
	
	\begin{equation}\label{formfactorfinal}
		G(\mathbf{k}n\mathbf{k}'n'\mathbf{q})=G_{\perp}F_zG_z
	\end{equation}
	where
	\begin{equation}\label{DiracDeltaTransverse}
		G_{\perp}=\frac{\delta(\mathbf{k}'_{\perp}-\mathbf{k}_{\perp}+\mathbf{q}_{\perp})}{N}
	\end{equation}
	
	\begin{widetext}
		\begin{equation}
			F_z(k_z,k'_z,q_z)=\frac{( (e^{\frac{1}{2} i (a (2 k'_z - 2 k_z N + q_z) + 2 b (k'_z - k_z N + q_z))}) (e^{i (a + b) k_z N} - e^{i (a + b) N (k'_z + q_z)})}{( (e^{i (a + b) k_z} - e^{i (a + b) (k'_z + q_z)})}
		\end{equation}
		\begin{equation}
			G_z(n,n',q_z)=\frac{(A_{nn'}-B_{nn'})}{(-\Sigma_{n,n'}\pi + a q_z) (-\Delta_{n,n'}\pi + a q_z) (\Delta_{n,n'}\pi + a q_z) (\Sigma_{n,n'} \pi + a q_z))}
		\end{equation}
		\begin{equation}
			A_{nn'}=2 n \pi \sin\left[\frac{n \pi}{2}\right] \left(-(1 + e^{i a q_z}) ((n'^2-n^2) \pi^2 + a^2 q_z^2) \cos\left[\frac{n' \pi}{2}\right] + 
			2 i a (-1 + e^{i a q_z}) n' \pi q_z \sin\left[\frac{n' \pi}{2}\right]\right)
		\end{equation}
		\begin{equation}
			B_{nn'}=  2 \cos\left[\frac{n \pi}{2}\right] \left(-i a (-1 + e^{i a q_z}) q_z ((n^2  + n'^2) \pi^2 -  a^2 q^2_z) \cos\left[\frac{n' \pi}{2}\right] + (1 + e^{i a q_z}) n' \pi ((n^2 - n'^2) \pi^2 +
			a^2 q_z^2) \sin\left[\frac{n' \pi}{2}\right]\right)
		\end{equation}
		where $\Sigma_{n,n'}=n+n',\Delta_{n,n'}=n'-n$. For intra-miniband scattering ($n'=n$), the expression for  $ G(\mathbf{k}n\mathbf{k}'n'\mathbf{q})$ simplifies to
		\[
		G(\mathbf{k}n\mathbf{k}'n'\mathbf{q})=\frac{\delta(\mathbf{k}'_{\perp}-\mathbf{k}_{\perp}+\mathbf{q}_{\perp})}{N}
		\]
		\begin{equation}
			\times
			\frac{-((i e^{\frac{1}{2} i (a (2 k'_z - 2 k_z N + q_z) + 2 b (k'_z - k_z N + q_z))}) (-1 + e^{
					i a q_z}) (e^{i (a + b) k_z N} - e^{i (a + b) N (k'_z + q_z)}) (-4 n^2 \pi^2 + (1 + (-1)^
				n) a^2 q_z^2)}{(
				(e^{i (a + b) k_z} - e^{i (a + b) (k'_z + q_z)}) (q_z a) (-2 n \pi + 
				a q_z) (2 n \pi + a q_z))}.
		\end{equation}
	\end{widetext}
	
	\begin{figure}
		\includegraphics[angle=0,origin=c, scale=0.5]{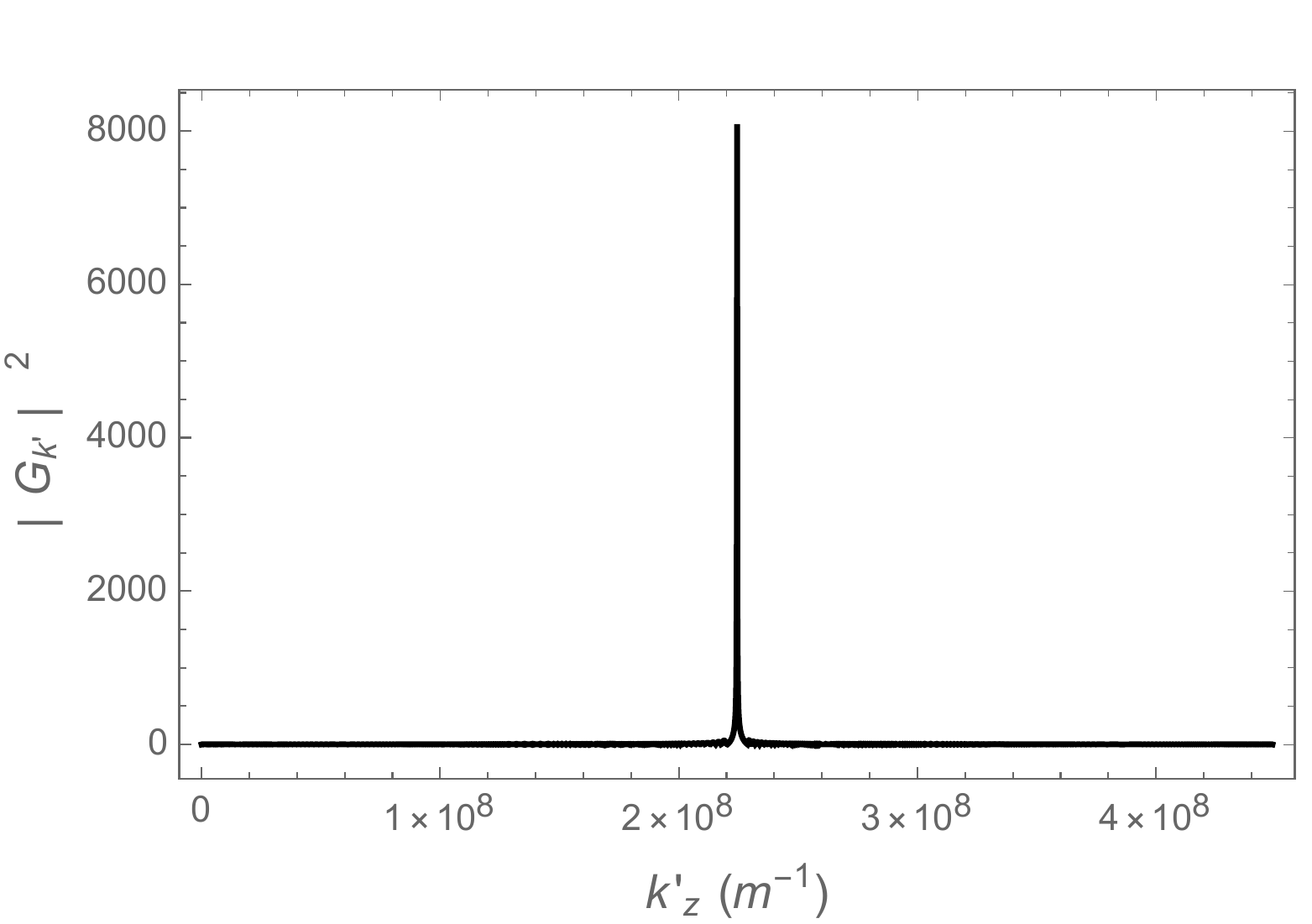}
		\label{fig:SuperlatticeFormFactor}
		\caption{
			The profile of the form factor of electron-phonon interaction in a superlattice consisting of wells of width $a$ and barriers of width $b$. The peak is located at $k'_z=k_z-q_z$.}
	\end{figure}
	
	The profile of $|G(\mathbf{k}n\mathbf{k}'n'\mathbf{q})|^2$ is numerically computed and illustrated in Fig. above in this Supplementary Materials, plotted as function of $k'_z$ while all other parameters fixed ($n = 1, n' = 1, N = 10^6, a = 5\mathrm{nm}, b = 2 \mathrm{nm}, k_z=4.48\times 10^8 m^{-1}, q_z=2.24\times 10^8 m^{-1}, k'_{\perp}=k_{\perp}=q_{\perp}=0$). It is clear that the profile takes the form of Dirac delta function, centered at $k'_z=k_z-q_z$. As such, combining with the Dirac delta function in Eq.(\ref{DiracDeltaTransverse}), we can write   
	\begin{equation}\label{DiracDeltaFormFactorS}
		|G(\mathbf{k}n\mathbf{k}'n'\mathbf{q})|^2=C_{N}(n,n',k_z,q_z)\delta(\mathbf{k}'-\mathbf{k}+\mathbf{q})
	\end{equation}
	where the coefficient $C_N(n,n', k_z, q_z)$ is to be determined from the normalization condition, giving
	\begin{equation}
		C_{N}(n,n',k_z,q_z)=\frac{2\pi}{(a+b)}\tilde{F}_N(k_z,q_z)|G_z(n,n',q_z)|^2.
	\end{equation}
	
	The real function $\tilde{F}_N(k_z,q_z)$ in the above equation is determined from the relation
	\begin{equation}
		\frac{1}{N^2}  \int^{\frac{\pi}{(a+b)}}_{-\frac{\pi}{(a+b)}} dk'_z |F_z(k_z,k'_z,q_z)|^2=\frac{2\pi}{(a+b)}\tilde{F}_N(k_z,q_z).
	\end{equation}
	It turns out that $\tilde{F}_N(k_z,q_z)$ converges to $1/N$; where $N$ is the number of periods of the superlattice.
	\comment{
		giving 
		\begin{equation}
			\tilde{F}(k_z,q_z)=-\frac{i}{2\pi}\mathcal{F}(k_z,q_z)
		\end{equation}
		where $\mathcal{F}(k_z,q_z)$ is a rather complicated mathematical expression involving Hypergeometric functions
		\begin{widetext}
			\[
			\mathcal{F}(k_z,q_z)=\frac{((2 e^{i(a + b)(-k_{UV} + q_z)} - e^{i(a + b) (k_{UV} (-1 + N) + k_z N + q_z - N q_z)})}{(
				e^{i (a + b) k_z} - e^{i(a + b) (-k_{UV} + q_z)})} + \frac{-2 e^{i(a + b)(k_{UV} + q_z)} + e^{
					i (a + b) (k_{UV} - k_{UV} N + k_z N + q_z - N q_z)}}{e^{i(a + b) k_z} - e^{i(a + b) (k_{UV} + q_z)}}
			\]
			\[
			+ \frac{
				e^{i(a + b)(-1 + N)(k_{UV} + k_z - q_z)}
				N \mathrm{Hypergeometric2F1}[1, 1 - N, 2 - N, 
				e^{-i (a + b) (k_{UV} + k_z - q_z)}]}{-1 + N} 
			\]
			\[
			- \frac{(
				e^{-i (a + b) (-1 + N) (k_{UV} - k_z + q_z)}
				N \mathrm{Hypergeometric2F1}[1, 1 - N, 2 - N, e^{
					i (a + b) (k_{UV} - k_z + q_z)}]}{-1 + N} 
			\]
			\[
			- \frac{
				e^{-i(a + b) (1 + N) (k_{UV} + k_z - q_z)} \mathrm{Hypergeometric2F1}[2, 1 + N, 2 + N, 
				e^{-i(a + b)(k_{UV} + k_z - q_z)}]}{1 + N}
			\]
			\begin{equation}
				+ \frac{(
					e^{i (a + b) (1 + N) (k_{UV} - k_z + q_z)}
					\mathrm{Hypergeometric2F1}[2, 1 + N, 2 + N, e^{
						i (a + b) (k_{UV} - k_z + q_z)}])}{(1 + N)}
			\end{equation}
		\end{widetext}
		which is supposed to be a purely imaginary-valued function of $k_z,q_z$ so as to give a real-valued $\tilde{F}(k_z,q_z)$.}
	
	\section{Derivation of Electron-Phonon Scattering Rate}
	
	The change in the electron energy (denoted by $\Delta E_{n\mathbf{k};n'\mathbf{k}',\mathbf{q}}$ in Eq. (15) in the main text) upon electron-phonon scattering reads
	\begin{widetext}
		\begin{equation}
			E_n(\mathbf{k}_{\perp},k_z)-E_{n'}(\mathbf{k}_{\perp}-\mathbf{q}_{\perp},k'_z)    =\frac{\hbar^2}{2m_c}\left[(k^2_{nz}-{k_{n'z}}^2)-q^2_{\perp}+2k_{\perp}q_{\perp}\cos\phi \right]+2V_{n'}\cos k'_z(a+b)-2V_{n}\cos k_z(a+b)
			\ ,
		\end{equation}
	\end{widetext}
	where 
	\begin{equation}
		E_n(a)=\frac{\hbar^2k^2_{nz}}{2m_c}
	\end{equation}
	and $\phi$ is the angle between ${\bf k_\perp}$ and ${\bf q}_\perp$. The distinction between $k_{nz}$ and $k_z$ is to be noted; the former gives the energy level of a confined electron state (the center of the mini-subband) while the latter is a quasi-momentum of the electron in the superlattice direction due to the periodicity of the structure. Within infinite potential well approximation, $k_{nz}=n\pi/a$ while within finite potential well picture using the depth of the well $V_c$ in Fig. 1 in the main text, it is the energy level $E_n(a)$ that is determined from the solution of transcendental equation from solving Schrodinger equation for a finite potential well. The following derivation uses infinite potential well approximation. The corresponding finite potential well results are obtained simply by the following replacement
	\begin{equation}
		k^2_{nz}=\frac{n^2\pi^2}{a^2}\rightarrow \frac{2m_cE_n(a)}{\hbar^2}
	\end{equation}
	to get the final results of this section.
	
	Using 
	\begin{equation}\label{transformationsumtointegral}
		\sum_{\mathbf{k}_{\perp}}\rightarrow \frac{S}{(2\pi)^2}\int_0^{2 \pi} d \phi \int_0^{+\infty} d{k}_{\perp} k_\perp \ ,
	\end{equation}
	we get
	\begin{widetext}
		\[
		A_{\mathbf{q}_{\perp},q_z}=\frac{2 m_c S}{(2\pi)^2\hbar^2 q_\perp}
		\sum_{n,n';k_z,k'_z}
		\int_0^{2 \pi} d \phi \int_0^{+\infty} d{k}_{\perp}
		I^2_{\mathbf{k}_{\perp},n,\mathbf{k}_{\perp}-\mathbf{q}_{\perp},n'}
		\left[(N_{\mathbf{q}_{\perp},q_z}+1)f_{\mathbf{k}_{\perp},n}(1-f_{\mathbf{k}_{\perp}-\mathbf{q}_{\perp},n'})
		-N_{\mathbf{q}_{\perp},q_z}f_{\mathbf{k}_{\perp}-\mathbf{q}_{\perp},n'}(1-f_{\mathbf{k}_{\perp},n})\right]
		\]
		\begin{equation}\label{rateofchangeofbosonnumberS}
			\times \delta\left( \cos \phi +\frac{(n^2-{n'}^2)}{2 k_\perp q_\perp}\frac{\pi^2}{a^2} - \frac{q_\perp}{2 k_\perp}  
			- \frac{m_c \omega_{q_{\perp},q_z} }{\hbar k_\perp q_\perp} + \frac{m_c}{\hbar^2k_{\perp}q_{\perp}}\left(2V_{n'}\cos k'_z(a+b)-2V_{n}\cos k_z(a+b)\right)
			\right)
		\end{equation}
	\end{widetext}
	where $\sum_{n,n'}$ represents the sum over mini-subbands while $I^2_{\mathbf{k}_{\perp},n,\mathbf{k}_{\perp}-\mathbf{q}_{\perp},n'}=|G(\mathbf{k}_{\perp},k_z,n;\mathbf{k}_{\perp}-\mathbf{q}_{\perp},k'_z,n';\mathbf{q}_{\perp},q_z)|^2|\tilde{W}_{q_z}|^2$. The argument of the Dirac delta function can be written as $\delta(\cos\phi-\cos\phi_0)$.
	
	Since $|\cos\phi|\leq 1$, we obtain a constraint on the quantity following the $\cos\phi$ within the Dirac delta function in Eq.(\ref{rateofchangeofbosonnumberS}) above, resulting in $k_{\perp}\geq k^{\mathrm{min}}_{\perp}(n,n')$ where
	\begin{widetext}
		\begin{equation}
			k^{\mathrm{min}}_{\perp}(n,n')= \frac{q_\perp}{2 } +\frac{({n'}^2-n^2)}{2  q_\perp}\frac{\pi^2}{a^2} 
			+ \frac{m_c \omega_{q_{\perp},q_z} }{\hbar  q_\perp} - \frac{m_c}{\hbar^2q_{\perp}}\left(2V_{n'}\cos k'_z(a+b)-2V_{n}\cos k_z(a+b)\right)
		\end{equation}
	\end{widetext}
	giving the minimum transverse wave vector that an electron must have in order to be able to emit an LO phonon. Integrating over $\phi$, we have 
	\begin{equation}
		\cos\phi_0= \pm \frac{k^{\mathrm{min}}_{\perp}(n,n')}{k_{\perp}}
	\end{equation}
	giving 
	\begin{equation}
		\int^{2\pi}_0 d\phi\delta(\cos\phi-\cos\phi_0)=\frac{2k_{\perp}}{\sqrt{k^2_{\perp}-(k^{\mathrm{min}}_{\perp}(n,n'))^2}}.
	\end{equation}
	Applying the above result to Eq.(\ref{rateofchangeofbosonnumberS}) gives
	\begin{widetext}
		\[
		A_{\mathbf{q}_{\perp},q_z}=\frac{2 m_cS}{(2\pi)^2\hbar^2 q_\perp} \sum_{n,n';k_z,k'_z}
		\int_{k^{\mathrm{min}}_{\perp}}^{+\infty} d{k}_{\perp}
		I^2_{\mathbf{k}_{\perp},n,\mathbf{k}_{\perp}-\mathbf{q}_{\perp},n'}
		\left[(N_{\mathbf{q}_{\perp},q_z}+1)f_{\mathbf{k}_{\perp},n}(1-f_{\mathbf{k}_{\perp}-\mathbf{q}_{\perp},n'})
		-N_{\mathbf{q}_{\perp},q_z}f_{\mathbf{k}_{\perp}-\mathbf{q}_{\perp},n'}(1-f_{\mathbf{k}_{\perp},n})\right]
		\]
		\begin{equation}\label{rateofchangeofbosonnumberS}
			\times\frac{2k_{\perp}}{\sqrt{k^2_{\perp}-\left(\frac{(n^2-{n'}^2)}{2  q_\perp}\frac{\pi^2}{a^2} - \frac{q_\perp}{2 }  
					- \frac{m_c \omega_{q_{\perp},q_z} }{\hbar  q_\perp} + \frac{m_c}{\hbar^2q_{\perp}}\left(2V_{n'}\cos k'_z(a+b)-2V_{n}\cos k_z(a+b)\right) \right)^2}}.
		\end{equation}
	\end{widetext}
	
	The final expression for the rate of change of number of LO phonons is given by
	\[
	\frac{dN_{\mathbf{q}_{\perp},q_z}}{dt}
	=( N_{\mathbf{q}_{\perp},m}(T_c) - N_{\mathbf{q}_{\perp},m})
	\]
	\[
	\times
	\left[2\frac{2\pi}{\hbar}\left(\frac{e^2\hbar\omega_{q_{\perp},m}}{q_{\perp} }
	\left[\frac{1}{\varepsilon_{\infty}}-\frac{1}{\varepsilon_s}\right]\frac{|\tilde{W}_{q_z}|^2}{8\pi q^2V}\right)\right]\frac{2m_c}{\hbar^2} \frac{S}{(2\pi)^2}
	\]
	\[
	\times\sum_{n,n';k_z,k'_z}|G_{n,n'}(k_z,k'_z,q_z)|^2\int^{+\infty}_{k^{\mathrm{min}}_{\perp}} dk_{\perp}\left[f_{\mathbf{k}_{\perp}-\mathbf{q}_{\perp},n'}-f_{\mathbf{k}_{\perp},n}\right] 
	\]
	\begin{equation}\label{rateofchangeofbosonnumberSfinal}
		\times   \frac{k_{\perp}}{\sqrt{k^2_{\perp}-(k^{\mathrm{min}}_{\perp})^2}},
	\end{equation}
	where we have made use of a nice identity \cite{Wurfel},
	\begin{equation}
		f_{\mathbf{k}_{\perp},n}(1-f_{\mathbf{k}_{\perp}-\mathbf{q}_{\perp},n'}) =N_{\mathbf{q}}(T_c)(f_{\mathbf{k}_{\perp}-\mathbf{q}_{\perp},n'}-f_{\mathbf{k}_{\perp},n}) \ ,
	\end{equation}
	where $N_{\mathbf{q}}(T_c)=1/(\exp(\hbar \omega_{LO}/(k_BT_c))-1)$ is the equilibrium phonon distribution at the carrier temperature $T_c$.
	
	To arrive at the final result given in Eq.(17) in the main text, we replace the sums over $k_z,k'_z$ with integrals over $k_z,k'_z$,
	\[
	\sum_{k_z}h(k_z)\rightarrow
	\]
	\begin{equation}
		\mathrm{lim}_{N\rightarrow \infty}    \sum^{\frac{N}{2}}_{j=-\frac{N}{2}}h(k_z(j))\rightarrow N(a+b)\int^{\frac{\pi}{(a+b)}}_{-\frac{\pi}{(a+b)}}\frac{dk_z}{2\pi}h(k_z)
	\end{equation}
	where $k_z(j)=2\pi j/(N(a+b))$ and similarly for $k'_z$. Evaluating the integral over $k'_z$ gives
	\begin{widetext}
		\[
		\frac{dN_{\mathbf{q}_{\perp},q_z}}{dt}
		=\frac{N_{\mathbf{q}}(T_c) - N_{\mathbf{q}}}{\tilde{\tau}^{c-LO}_{\mathbf{q}}}=(N_{\mathbf{q}}(T_c) - N_{\mathbf{q}})\left[2N\frac{2\pi}{\hbar}\left(\frac{e^2\hbar\omega_{\mathbf{q}}}{q_{\perp} }\left[\frac{1}{\varepsilon_{\infty}}-\frac{1}{\varepsilon_s}\right]\frac{|\tilde{W}_{q_z}|^2}{8\pi q^2 V}\right)\right]\frac{2m_c}{\hbar^2}\frac{S}{(2\pi)^2}\sum_{n,n'}\int^{\frac{\pi}{(a+b)}}_{-\frac{\pi}{(a+b)}}\frac{dk_z}{(2\pi)^2}
		\]
		\begin{equation}\label{rateofchangeofbosonnumberS}
			\times \Theta\left(\frac{\pi}{(a+b)}-k_z+q_z\right)\Theta\left(k_z-q_z+\frac{\pi}{(a+b)}\right)C_N(n,n',k_z, q_z)(a+b)^2\frac 1 2 \frac{\sqrt{2 m_c k_B T_c}}{\hbar} \Gamma \Bigl( \frac 1 2\Bigr ) (Li_{1/2}(- e^{x_2}) -Li_{1/2}(-e^{x_1}) )
		\end{equation}
	\end{widetext}
	where $N_{\mathbf{q}}(T_c)=(\exp(\hbar\omega_{\mathbf{q}}/(k_BT_c))-1)^{-1}$, $\Theta(\cdots)$ is Heaviside theta function, $\Gamma$ is gamma function and $Li_{\alpha}(x)$ is polylogarithmic function of order $\alpha$ while $x_1,x_2,k^{\mathrm{min}}_{\perp}$ are given in the following paragraph. It is to be noted that the limits of integration over $k_z$ imposed by the product of two Heaviside functions in Eq.(\ref{rateofchangeofbosonnumberS}) indicates the inclusion of Umklapp scattering of electron by LO phonon for any finite wave vector $q_z$ of the latter. We have included the phonon spatial coherence weight factor squared $|\tilde{W}_{q_z}|^2$ that is discussed in the following section. 
	
	To obtain the last line of Eq. (\ref{rateofchangeofbosonnumberS}), we rewrite and evaluate the integral in Eq.(\ref{rateofchangeofbosonnumberSfinal}) as follows,
	\begin{equation}
		I =  \int^{\infty}_{k^{\mathrm{min}}_{\perp}} dk_{\perp}\frac{k_{\perp}}{\sqrt{k^2_{\perp}-(k^{\mathrm{min}}_{\perp})^2}}
		(f_1-f_2) \ ,
	\end{equation}
	where $f_i = \frac{1}{e^{\tilde{x}_i}+1}$, with respectively 
	\begin{equation}
		\tilde{x}_1= \beta (E_{n'}(k_{\perp})-2V_{n'}\cos (k_z-q_z) (a+b) -\mu_c)-\beta \hbar \omega_{LO}
	\end{equation}
	and 
	\begin{equation}
		\tilde{x}_2 = \beta (E_n(k_{\perp})-2V_n\cos k_z (a+b) -\mu_c),
	\end{equation}
	with $\mu_c$ the carrier chemical potential and 
	$\beta = 1/k_B T_c$. An obvious change of variable leads to
	\begin{equation}\label{IntegralI}
		I = \frac 1 2 \frac{\sqrt{2 m_c k_B T_c}}{\hbar}  \int_0^{+\infty} dx \frac{1}{\sqrt{x}} \Bigl[ \frac{1}{e^{(x-x_1)}+1} -\frac{1}{e^{(x-x_2)}+1}  \Bigr] \ ,
	\end{equation}
	with
	\[
	x_1 =-\beta \Bigl[\frac{\hbar^2 }{2 m_c}(k^{min}_\perp)^2+E_{n'}(a) -2V_{n'}\cos (k_z-q_z)(a+b)\Big]
	\]
	\begin{equation}\label{x1}
		+\beta (\mu_c+\hbar \omega_{LO}),
	\end{equation}
	\begin{equation}\label{x2}
		x_2= -\beta \Bigl[\frac{\hbar^2 }{2 m_c}(k^{min}_\perp)^2+E_n(a) -2V_{n}\cos k_z(a+b) -\mu_c\Big]
	\end{equation}
	where $k^{min}_{\perp}$ is the minimum transverse wave vector an electron must posses to be able to emit an LO phonon, 
	\[
	k^{\mathrm{min}}_{\perp}(n,n')= \frac{q_\perp}{2 } +\frac{(E_{n'}(a)-E_n(a))}{2 q_\perp}\frac{2m_c}{\hbar^2} 
	+ \frac{m_c \omega_{q_{\perp},q_z} }{\hbar  q_\perp}
	\]
	\begin{equation}
		- \frac{m_c}{\hbar^2q_{\perp}}\left(2V_{n'}\cos k'_z(a+b)-2V_{n}\cos k_z(a+b)\right)
	\end{equation}
	and $\mu_c=\mu_e$ the electron chemical potential.
	Equation (\ref{IntegralI}) can be expressed in terms of a polylogarithmic function using Ref.~\onlinecite{nist},
	\begin{equation}
		\int_0^{+\infty} \frac{dt}{\sqrt{t}} \frac{1}{e^{t-u}+1}= - \Gamma \Bigl(\frac 1 2 \Bigr) Li_{1/2} (- e^u) \ .
	\end{equation}
	This leads to
	\begin{equation}
		I =  \frac 1 2 \frac{\sqrt{2 m_c k_B T_c}}{\hbar} \Gamma \Bigl( \frac 1 2\Bigr ) (Li_{1/2}(- e^{x_2}) -Li_{1/2}(-e^{x_1}) ) \, ,
	\end{equation}
	which gives Eq.(\ref{rateofchangeofbosonnumberS}) when substituted into Eq.(\ref{rateofchangeofbosonnumberSfinal}).

	\begin{table*}[t]
		\begin{tabular}{|p{10cm}||p{2cm}|p{4cm}|}
			\hline
			\multicolumn{3}{|c|}{Material Parameters (InAs(well) AlAsSb(barrier))}\\
			\hline
			Quantity &  Symbol & Value \\
			\hline
			LO phonon energy & $E_0$ & $30$ meV \\
			Conduction band well depth & $V_c$ & 1.92 eV\\
			Energy gap (InAs) & $E_g$ & 0.354 eV\\
			Valence band well depth & $V_h$ & 0.05 eV\\
			Energy gap (AlAsSb) & $E_g$ & 2.22 eV\\
			Electron effective mass (InAs) & $m_c$ & 0.023 $m_0$ \\
			Heavy hole effective mass (InAs) & $m_{hh}$ & 0.410 $m_0$ \\
			Electron effective mass (AlAsSb) & $m_c$ & 0.1369 $m_0$ \\
			Heavy hole effective mass (AlAsSb) & $m_{hh}$ & 0.98 $m_0$ \\
			\hline
			Sound velocity (InAs) & $v_s$ & $4.4\times 10^3$ m/s \\
			Infinite-frequency susceptibility (InAs) & $\varepsilon_{\infty}$ & 15.15\\
			Static susceptibility (InAs) & $\varepsilon_s$ & 12.30\\
			Gruneisen parameter (InAs) &  $\gamma$ & 0.28\cite{SparksSwenson}
			\\
			Zero temperature LO phonon decay time (InAs) &  $\tau^{LO-ac}_0$ & $4.363\times 10^{-10}$ s
			\\
			Lattice constant (InAs) &  $a$ & 6.06 \AA
			\\
			Mass density (InAs) &  $\rho$ & 5680 kg/m$^3$\\ 
			\hline
		\end{tabular}
		\caption{The parameters of the semiconductor materials for the superlattice used in our calculations. The masses are expressed in terms of free electron mass $m_0=9.11\times 10^{-31}$kg.}
		\label{tbl:parameters}
	\end{table*}
	
	\section{Material and Operational Parameters}
	In this section of the Supplementary Materials, we provide the material and operational parameters that we used to produce the figures given in the main text. The superlattice materials are taken to be InAs for the well layer and AlAsSb for the barrier layer.\comment{We assume a fixed chemical potential $\mu_=0$eV to compute the thermalization power, independent of well and layer well thickness and used for both 3D (that is bulk) theory\cite{TsaiS} and our 2D theory of superlattice.} The corresponding carrier (electron) can be computed from the density of states given in the Section I of the Supplementar Materials. For the calculation of electron thermalization power of interest, only the parameters of electron in the well layer (InAs) enter directly in the equations in the main text. The carrier parameters for the holes in the well layer (InAs) and those of the electrons and the holes for the barrier layer (AlAsSb) however enter indirectly in the determination of the tunneling energy $V_n$ appearing in Eq.(1) in the main text, and so they are listed as well in Table I. The zero temperature LO phonon decay time is determined from using \cite{MakhfudzJPDAP}
	\begin{equation}\label{eq:tauLO-ac_0}
		\tau^{LO-ac}_0=\frac{32\pi\rho v^3_s}{\Gamma^2\hbar\omega^3_{\mathrm{LO}}}.
	\end{equation}
	where \cite{RidleyBookS}
	\begin{equation}\label{eq:gamma}
		\Gamma=\sqrt{\frac{4}{3}}\frac{\gamma\omega_{\mathrm{LO}}}{v_s} \ ,
	\end{equation}
	with $E_0=\hbar\omega_{\mathrm{LO}}$. In addition to the material parameters given in the Table, we have to assume the numerical values for the  parameters representing the operating condition. These include the lattice temperature $T_L$, carrier temperature $T_c$ and the quasi-Fermi level splitting $\Delta \mu = \mu_e - \mu_h$  where $\mu_e(\mu_h)$ are respectively the electron (hole) chemical potential.
	
	Normally, $T_L$ can be fixed to be an appropriate ambient temperature. We take $T_L=300$K to describe solar cell operation at ambient room temperature. On the other hand, the carrier parameters $T_c$ and $\Delta\mu$ are ought to be either determined from a detailed balance equation involving a given absorption of radiation and recombination process, or be taken from available experimental data. In this work, we do not pursue the former, but adopt the later instead, partially, where we fix $T_c$ and assume some appropriate values for $\Delta\mu$, which is empirically constrained to be $0\leq \Delta\mu\leq E_g$, where $\Delta\mu=0$ corresponds to absence of irradiation, with $E_g$ the band gap of the well layer. In producing Fig. 3 in the main text, we have taken $T_c=450$K, and $\Delta \mu = 0.354$eV, taken to be equal to $E_g$ for InAs, which are relatively high $T_c$ and large $\Delta\mu$ but are generally consistent with empirical observation that $T_c$ is higher when $\Delta\mu$ is larger. In producing Fig. 4 in previous section on the other hand, to make direct comparison with the data presented in \cite{EsmaielpourAPLs}, we have set $T_c=345$K and assumed $\Delta\mu=0.2$eV, while setting $a=2$nm. The fact that $P_{\mathrm{th}}$ in our result is very small (in the order of 10$^{-13}$W/cm$^2$) for such small thickness, since only one mini-subband exists while phonon-driven electron thermalization mainly comes from inter-mini-subband scattering, poses no problem because our theory strictly speaking assumes an infinitely long $N\rightarrow\infty$ superlattice; the total thermalization power will be finite as one considers $N$ times the $P_{\mathrm{th}}$ per period of superlattice that we compute. It is for this reason that in Fig. 4, we only present the normalized $P_{\mathrm{th}}$ relative to its large barrier thickness limit $P_{\mathrm{th0}}$ as the quantity that is appropriately compared with experiment \cite{EsmaielpourAPLs}.
	
	For carrier photogeneration by monochromatic irradiation (with laser, for example), given the thermalization power, the necessary absorbed power (with incoherent phonons) can be estimated using the following relation \cite{GiteauJAPs}
	\begin{equation}
		P_{\mathrm{th}}=\frac{E_{\mathrm{laser}}-(E_g+3k_BT_L)}{E_{\mathrm{laser}}}P_{\mathrm{abs}}
	\end{equation}
	where $E_{\mathbf{laser}}=hc/\lambda$ where $h$ is Planck constant, $c$ the speed of light, and $\lambda$ the (monochromatic) laser wave length. Taking $\lambda=532$nm, the absorbed power of about $P_{\mathrm{abs}}=14.8$ W/cm$^2$ for the upper subfigure of Fig. 3 in the main text, as an example.

\end{document}